\documentclass[reprint, aps, pra, onecolumn, floatfix, longbibliography, superscriptaddress, nofootinbib]{revtex4-2}
\usepackage{amsmath, amssymb}
\usepackage{bm}
\usepackage{booktabs}
\usepackage{graphicx}
\usepackage[usenames, dvipsnames]{color}
\usepackage{mathtools}
\usepackage{hyperref}
\hypersetup{colorlinks=true, linkcolor = blue, citecolor = blue, urlcolor = blue}
\usepackage{orcidlink}
\usepackage{xcolor}
\usepackage{bbold}
\usepackage{pifont}
\newcommand{\Lagr}{\mathcal{L}}

\newcommand{\dd}{\mathrm{d}}
\newcommand{\ee}{\mathrm{e}}
\newcommand{\ii}{\mathrm{i}}

\newcommand{\pdt}{\frac{\partial}{\partial t}}

\newcommand{\res}{\mathrm{r}}    
\newcommand{\wg}{\mathrm{w}}     
\newcommand{\qb}{\mathrm{q}}     
\newcommand{\vac}{\mathrm{vac}}  
\newcommand{\Fent}{F} 
\newcommand{\BellState}{\mathtt{B}}
\newcommand{\JJ}{\mathrm{J}}    
\newcommand{\vacket}{|\vac\rangle}

\begin{document}
\title{On-demand frequency-entangled microwave biphotons from a transmon--resonator dimer}
\author{E.~V.~Stolyarov\,\orcidlink{0000-0001-7531-5953}}
\email{eugene.stolyarov@bitp.kyiv.ua} 
\affiliation{Quantum Optics and Quantum Information Group, Bogolyubov Institute for Theoretical Physics, National Academy of Sciences of Ukraine, Vulytsya Metrolohichna 14-b, Kyiv 03143, Ukraine}
\author{O.~V.~Kliushnichenko\,\orcidlink{0000-0003-4806-8971}}
\email{kliushnychenko@iop.kiev.ua}
\affiliation{Institute of Physics, National Academy of Sciences of Ukraine, Prospect Nauky 46, Kyiv 03028, Ukraine}
\affiliation{Quantum Optics and Quantum Information Group, Bogolyubov Institute for Theoretical Physics, National Academy of Sciences of Ukraine, Vulytsya Metrolohichna 14-b, Kyiv 03143, Ukraine}
\begin{abstract}

We propose a superconducting circuit QED source of on-demand frequency-entangled microwave photon pairs, with one photon emitted into each of two separate waveguides, enabling individual manipulation.
The design is based on a transmon--resonator dimer, which consists of two transmon--resonator subsystems coupled through a cross-Kerr interaction.
We outline a viable circuit-QED realization in which the required cross-Kerr interaction is implemented using Josephson-junction couplers.
To analyze the emission dynamics, we derive and numerically solve the set of evolution equations that enable tracking the evolution of the joint spectral intensity of photons.
We characterize the resulting photon-number correlations, frequency anticorrelation, and spectral entanglement.
By varying the coupling strengths, we obtain biphotons with distinct joint spectral intensities and Schmidt spectra, ranging from broadly distributed multimode entanglement to states dominated by two nearly equally weighted Schmidt contributions.
In the latter regime, optimizing the transmon--resonator coupling and detuning yields an output biphoton state that has fidelity $F \approx 0.956$ with a two-dimensional maximally entangled state.
Restricting the local target modes to separated frequency bins yields a fidelity \(F_{\mathrm{bin}}\approx0.949\), indicating that the output state closely approximates a frequency-bin Bell pair.
Thus, the proposed microwave biphoton source provides a route toward distributing frequency-encoded quantum resources between remote superconducting nodes.

\end{abstract}
\setcounter{page}{1}
\maketitle
\section{Introduction}
\label{sec:Intro}

Photons provide a natural interface between spatially separated quantum systems, serving as flying qubits that transfer quantum information between stationary quantum memories and quantum processing units~\cite{kimble2008, ritter2012, northup2014}.
Hence, the controlled generation and manipulation of photonic quantum states are essential for quantum information technologies~\cite{slussarenko2019apr, flamini2019rpp}.

In particular, entangled states of photons constitute an important resource in quantum science.
These quantum states are harnessed in a wide range of established as well as emerging quantum technologies, such as photonic quantum computation \cite{kok2007}, quantum metrology and sensing \cite{giovannetti2011}, implementation of quantum communication protocols \cite{gisin2007}, quantum teleportation \cite{bouwmeester1997}, and quantum key distribution \cite{xu2020rmp}.
They also enable experimental tests of the foundations of quantum mechanics~\cite{aspect1981, howell2004prl, giustina2015}.
These applications motivate the development of efficient protocols for the generation of entangled photonic states.

In the microwave domain, circuit quantum electrodynamics (QED) offers a platform in which superconducting artificial atoms interact with quantized electromagnetic fields~\cite{blais2004, blais2021}.
This versatile platform is a leading hardware architecture for the implementation of quantum information processing devices.
Microwave quanta can transfer quantum information between remote superconducting chips~\cite{zhong2021nature, niu2023natelectr}. 
Controlled emission and absorption of microwave photons have enabled deterministic quantum-state transfer and entanglement generation between spatially separated superconducting qubits~\cite{kurpiers2018nat, zhong2021nature}.
Sources of entangled propagating microwave photons would complement these interfaces by providing shared quantum resources to distinct nodes.
A particularly useful design involves generating photon pairs on demand and directing individual photons into separate waveguides or transmission lines.
Such a setup would enable independent transmission, manipulation, and absorption of each photon in a pair.
For example, measurement of one photon in a coherent frequency basis could enable remote spectral shaping of the other photon~\cite{sych2017pra}.

Several approaches to generating entangled microwave radiation have been demonstrated.
Nondegenerate Josephson mixers generate two-mode squeezed fields at different frequencies in separate transmission lines~\cite{flurin2012}, while voltage-biased Josephson junctions coupled to microwave resonators emit continuous entangled microwave beams through inelastic Cooper-pair tunneling~\cite{lepp2012, westig2017, peugeot2021}.
Alternatively, one can leverage nonlinearity of superconducting artificial atoms~\cite{clarke2008, kjaer2020}.
The anharmonic structure of their energy levels enables cascaded relaxation leading to emission of frequency-entangled photons~\cite{gaspar2017}.

Frequency entanglement is particularly attractive in the circuit QED platform as it encodes quantum correlations in a degree of freedom directly accessible through frequency-selective microwave components and coherent frequency conversion.
For a pure biphoton state with one photon in each output channel, frequency entanglement corresponds to the nonseparability of the joint spectral amplitude.
Its generation in separate microwave waveguides was considered in Ref.~\cite{stolyarov2022}, which proposed a circuit-QED setup based on the correlated relaxation of two initially excited emitters. In that scheme, a three-level emitter couples to both resonators and controls the relaxation of a second emitter through its state-dependent interaction with one of the resonators.

A different mechanism for the generation of frequency-entangled photon pairs was proposed by Governale \textit{et al.}~\cite{governale2025prxq}.
They considered a Cooper pair splitter comprising two double quantum dots coupled to separate microwave transmission lines.
Coherent electronic processes in that system produce photon pairs with frequency entanglement between the two output channels.

The aforementioned proposals motivate further studies of alternative schemes for the deterministic generation of microwave biphotons in superconducting circuits.
In this work, we present a viable design of an on-demand source of frequency-entangled microwave photon pairs, with each photon propagating in a separate waveguide.
In the next section, we elaborate on our design, explain the underlying principle of operation, and discuss its advantages.

\section{Operational principle}
\label{sec:Principle}

 \begin{figure}[t!] 
	\centering
	\includegraphics{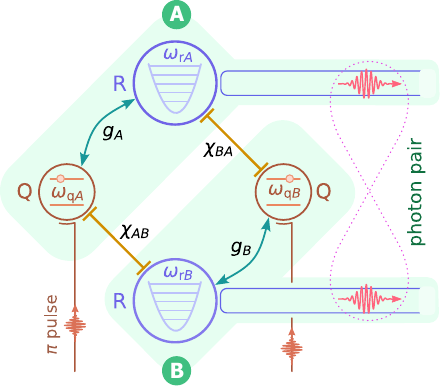}
	\caption{Scheme of the considered setup: The setup consists of two subsystems ($A$ and $B$).
    Each subsystem is comprised of a qubit linearly coupled to a resonator, which is coupled to a waveguide.
    Within the subsystem $s \in \{A,B\}$, the qubit and the resonator exchange an excitation with rate $g_s$.
    Subsystems interact via the cross-Kerr coupling: the qubit in one subsystem is coupled to the resonator in the other subsystem.
    Both qubits are prepared in their excited states by applying $\pi$ pulses~\cite{allen1975} delivered via individual control lines.
    \label{fig:Scheme}}
\end{figure}

Let us outline the general scheme underlying the operation of the setup studied.
Consider a system schematically illustrated in Fig.~\ref{fig:Scheme} that consists of two subsystems labeled $A$ and $B$.
Each subsystem is comprised of a two-level system (qubit), a single-mode resonator, and a waveguide.
The qubit and the resonators are coupled via linear interaction, which leads to coherent excitation exchange between them.
The resonator mode is also coupled to a waveguide, allowing photons to leak out of the subsystem.
Additionally, each qubit interacts via cross-Kerr coupling with the resonator belonging to the other subsystem.
The cross-Kerr interaction leads to state-dependent shifts in the frequencies of the coupled systems, yet without excitation exchange between them.
Thus, there is no excitation exchange between the subsystems $A$ and $B$.
However, cross-Kerr coupling ensures that the frequency of the resonator in the subsystem $A$ ($B$) depends on the state of the qubit in the subsystem $B$ ($A$) and vice versa.

Assume now that the qubits are prepared in their excited states.
Due to coupling to the leaky resonator, the excited qubit eventually relaxes, leading to the emission of a single photon into the waveguide to which the resonator is coupled.
Without cross-Kerr coupling, the two subsystems evolve independently.
Hence, the emitted photons exhibit neither temporal nor frequency correlations.
However, relaxation of the qubits is no longer uncorrelated with cross-Kerr coupling between the subsystems $A$ and $B$.
Indeed, the excited qubit in one subsystem detunes the resonator from the qubit in the other subsystem, affecting the excitation exchange between them.
Consequently, the emission processes in the two subsystems become correlated.
As demonstrated in Sec.~\ref{sec:CrossCorrelations}, the cross-Kerr interaction produces correlations between the waveguide populations during emission.
It also modifies the joint spectral amplitude (JSA) of the emitted photons, giving rise to frequency anticorrelation examined in Sec.~\ref{sec:FrequencyCorrelation}.

The scheme considered here replaces the multilevel, shared-resonator coupling arrangement of Ref.~\cite{stolyarov2022} with a direct cross-Kerr interaction between the qubit and the resonator in different subsystems.
The advantage of this approach stems from the fact that no frequency matching between the subsystems is required.
While the scheme in Ref.~\cite{stolyarov2022} requires selective resonance conditions between emitter transitions and resonator modes, the cross-Kerr coupling induces population-dependent frequency shifts without excitation exchange between the subsystems and remains operative regardless of their frequency detuning.
This provides more flexibility in choosing the operating frequencies of the biphoton source.

Moreover, as we demonstrate in Sec.~\ref{sec:OutputState} and discuss in detail in Sec.~\ref{sec:FrequencyBinEntanglement}, for a specific combination of the system parameters, the considered scheme enables generation of frequency-entangled photon pairs exhibiting narrow, well-separated spectral features.
This regime connects the continuous spectral structure of the biphoton to discrete frequency-bin encoding~\cite{lu2023opt}.

In the next section, we outline a circuit QED implementation of the general scheme discussed above. 
The rest of the manuscript is organized as follows.
We write down the model Hamiltonian describing the considered setup and discuss the approximations we made in Sec.~\ref{sec:Model}.
In Sec.~\ref{sec:Evolution}, we derive a set of equations of motion governing the evolution of a quantum state of the system and analyze the emission dynamics.
In Sec.~\ref{sec:OutputState}, we analyze the effect of the setup parameters on the entanglement degree of the emitted photons.
In Sec.~\ref{sec:FrequencyBinEntanglement}, we study the regime of generation of biphotons closely approximating frequency-bin Bell states.
We summarize our results and outline possible applications of the proposed biphoton source in Sec.~\ref{sec:Summary}.
Additional derivations and technical details are delegated to Appendixes.

\section{Implementation} \label{sec:Implementation}

The possible circuit QED realization of the system discussed in Sec.~\ref{sec:Principle} is as follows.
The two-level atoms are represented by transmons~\cite{koch2007, schreier2008prb}.
This type of superconducting artificial atom offers a simple design and a long coherence time that can reach hundreds~\cite{place2021, wang2022npj, bal2024njp, tuokkola2025ncom, joshi2026arxiv} and even thousands~\cite{bland2025nat} of microseconds.
That makes a transmon one of the most widely used artificial atom designs in circuit QED.
The states of the transmons can be prepared and manipulated with high accuracy by applying classical pulses delivered via individual control lines.
Alternatively, one can use single-flux quantum digital control~\cite{mcdermott2014prappl, *liebermann2016prappl, leonard2019prappl, liu2023prxq}.

The resonators are implemented by lumped-element $LC$ oscillators, as in Refs.~\cite{collodo2019, vrajitoarea2020}.
The advantage of this type of microwave resonators is genuine single-mode behavior.
In addition, lumped-element oscillators feature a small physical footprint and can offer high quality factors~\cite{lilishi2022, crowley2023, dhundhwal2025apl}.

Within the same subsystem, the transmons are coupled to the $LC$ oscillators via capacitive coupling, which leads to a linear interaction between them.
The transmons and the resonators belonging to different subsystems are coupled by Josephson junctions (JJs)~\cite{vrajitoarea2020, wang2025prl, beaulieu2026prxq} shunted by capacitances.
Inherent nonlinearity of a JJ coupler gives rise to a \emph{nonperturbative} cross-Kerr interaction between the transmon and the resonator.

Alternatively, instead of single capactitively-shunted JJs, one can use capactively-shunted dc SQUIDs as nonlinear couplers~\cite{koun2018, stolyarov2025detector, stolyarov2025prb, stolyarov2026pra, kim2026arxiv} bridging transmons and resonators. 
This enables \emph{in situ} tunability of cross-Kerr couplings by tuning the external fluxes via SQUID loops.
Each SQUID in the system can be controlled individually via a dedicated flux-bias line.

In addition to the cross-Kerr interaction, the coupler also induces a linear interaction between the transmon and the resonator in different subsystems.
If their frequencies are close, this coupling leads to coherent single-excitation exchange between different subsystems, which is detrimental to the operation of our device. 
However, this coupling can be cancelled by a specific choice of a capacitive shunt added in parallel to the coupler JJ or SQUID; see, e.g., Refs.~\cite{neumeier2013, ciani2019qst, sank2025prappl, stolyarov2026pra}.

In our scheme, we assume that the transmon and the resonator belonging to the same subsystem operate in the resonant or near-resonant regime, allowing for the most efficient excitation exchange between them.
However, in practice, fabrication imperfections can deteriorate the necessary frequency matching.
Hence, one may require tunability of frequencies to mitigate the effect of imperfections and achieve the resonant regime.
To implement frequency tuning in the considered system, one can use split transmons~\cite{barends2013} with single JJs replaced by dc SQUIDs.
Controlling the external fluxes via the SQUID allows one to adjust the frequency \emph{in situ}.
Additionally, introducing asymmetry in the transmon dc SQUID reduces sensitivity to flux noise \cite{koch2007, krantz2019}.
Alternatively, one can build transmons using voltage-tunable JJs~\cite{casparis2018, riechert2025, aguado2020}.

\section{Model}
\label{sec:Model}

\subsection{The Hamiltonian}
\label{sec:Hamiltonian}

The Hamiltonian that models the circuit QED system outlined above reads as
\begin{equation} \label{eq:FullHamiltonian}
 \hat H = \hat H_{\qb\res} + \hat H_\wg.
\end{equation}
Here, the first term describes a transmon--resonator dimer---a pair of transmon--resonator systems coupled by the cross-Kerr interaction.
The second term describes the waveguides into which the photons are emitted and their coupling to the transmon--resonator system.

The Hamiltonian describing the transmon--resonator system reads as follows (see detailed derivation in Appendix~\ref{app:CrossKerr})
\begin{equation} \label{eq:TransmonsHamiltonian}
  \begin{split} 
    \frac{\hat H_{\qb\res}}{\hbar} =
    \sum_{s} \omega_{\res s} \, \hat a^\dag_s \hat a_s 
    + \sum_{s} \omega_{\qb s} \, \hat \sigma^+_{s} \hat \sigma^-_{s} 
    + \sum_{s} g_s (\hat \sigma^+_{s} \hat a_s + \hat a^\dag_{s} \hat \sigma^-_{s})
    - \sum_{s} \chi_{s\bar{s}} \hat\sigma^+_s \hat \sigma^-_s \hat a^\dag_{\bar{s}} \hat a_{\bar{s}},
  \end{split}
\end{equation}
Here and throughout the rest of the paper, $s \in \{A,B\}$ always denotes a subsystem index, and the summation goes over all indices.
The first two terms constitute the diagonal part of the Hamiltonian.
The first term describes the resonators modeled as bosonic modes, where $\omega_{\res s}$ stands for the frequency of the resonator in subsystem $s$. 
Operator $\hat a_s$ annihilates an excitation in the resonator $s$.
The second term describes the qubits (transmons) modeled as two-level systems with $\omega_{\qb s}$ being the transition frequency of the transmon $s$. 
Operators $\hat \sigma^+_{s} \equiv |e\rangle_{s}\langle g|$ and $\hat \sigma^-_{s} \equiv |g\rangle_{s}\langle e|$ are the raising and lowering operators of the transmon $s$, where $|g\rangle_{s}$ and $|e\rangle_{s}$ denote its ground and excited states, respectively.

The last two terms in Eq.~\eqref{eq:TransmonsHamiltonian} describe interactions between the transmons and the resonators.
The penultimate term describes the coherent single-excitation exchange with rate $g_s$ between the transmon and the resonator mode belonging to the same subsystem $s$.
The last term in the Hamiltonian~\eqref{eq:TransmonsHamiltonian} describes the cross-Kerr interaction of strength $\chi_{s\bar{s}}$ between the transmon $s$ and the resonator in a different subsystem $\bar{s}$.
We define $\chi_{s\bar s}>0$ as the cross-Kerr coupling strength and include its negative sign explicitly in the Hamiltonian, in consistency with the derivation in Appendix~\ref{app:CrossKerr}.
Hereinafter, for convenience, index $\bar{s}$ always denotes a subsystem other than $s$, i.e., $\bar{s} = \{A, B\}\setminus\{s\}$.

The second term in the total system Hamiltonian~\eqref{eq:FullHamiltonian} reads as
\begin{equation} \label{eq:WaveguideHamiltonian}
 \begin{split} 
    \frac{\hat H_\wg}{\hbar} = \sum_{s} \int \dd\omega \, \omega \, \hat w^\dag_{s \omega} \hat w_{s \omega} 
    + \sum_{s} f_{s} \int \dd\omega (\hat a^\dag_{s} \hat w_{s \omega} + \hat w^\dag_{s \omega} \hat a_{s}). 
 \end{split}
\end{equation}
The upper line describes the waveguides, which are modeled as ensembles of noninteracting bosonic modes.
The operator $\hat w_{s \omega}$ annihilates a photon with frequency $\omega$ in a waveguide mode $s$ and satisfies the equal-time commutation relation $[\hat w_{s\omega}, \hat w^\dag_{s'\omega'}] = \delta_{ss'}\delta(\omega-\omega')$.
The bottom line in the Hamiltonian~\eqref{eq:WaveguideHamiltonian} describes linear coupling of strength $f_{s}$ between the resonator $s$ and the corresponding waveguide.
This interaction results in the spontaneous emission of a resonator photon into the waveguide with rate $\kappa_s = 2\pi f_s^2$.

\subsection{Approximations} \label{sec:Approximations}

\subsubsection{Rotating-wave and narrow-bandwidth approximations} \label{sec:ApproxRWA}
In the transmon--resonator Hamiltonian~\eqref{eq:TransmonsHamiltonian}, we applied the rotating-wave approximation (RWA) and neglected the fast-oscillating terms $\propto (\hat\sigma^+_{s} \hat a^\dag_{s} + \hat\sigma^-_{s} \hat a_{s})$.
This approximation is valid provided that the coupling strengths between the transmon and the resonator are much smaller than their frequencies $|g_{s}| \ll \omega_{\mathrm{r}s}, \omega_{\mathrm{q}s}$.
Additionally, for the RWA to be valid, the magnitude of the detuning between the resonator and qubit frequencies should be small compared to their sum $|\Delta_s| \ll \omega_{\mathrm{r}s}+\omega_{\mathrm{q}s}$, where $\Delta_s = \omega_{\mathrm{q}s}-\omega_{\mathrm{r}s}$ denotes the detuning.

In the interaction term of the Hamiltonian~\eqref{eq:WaveguideHamiltonian}, we do not account for the frequency dependence of the resonator-waveguide couplings.
In the intended operating regime, the microwave carrier frequencies greatly exceed the linewidths, coupling-induced splittings, and detunings that determine the emission spectrum.
Away from waveguide cutoffs~\cite{parrarodriguez2018,shitara2021}, the capacitive coupling strength $f_s(\omega)\propto\sqrt{\omega}$ (see, for example, Ref.~\cite{blais2021}) therefore varies negligibly across the entire emission band.
This justifies the frequency-independent approximation $f_s(\omega)\simeq f_s(\omega_{\res s})\equiv f_s$ used in Eq.~\eqref{eq:WaveguideHamiltonian}.
The weak resonator--waveguide coupling, $\kappa_s \ll \omega_{\res s}$, also justifies the RWA, in which we neglect the counter-rotating terms $\propto (\hat a_s^\dag \hat w_{s\omega}^\dag + \hat a_s\hat w_{s\omega})$.
Following the standard narrow-band Markov approximation of input--output theory~\cite{gardiner1985,combes2017apx}, we extend the frequency integrals in Eq.~\eqref{eq:WaveguideHamiltonian} from $[0,+\infty)$ to $(-\infty,+\infty)$.
The added negative-frequency region lies far outside the relevant emission band and carries negligible spectral weight.
Throughout the manuscript, integration limits over photon frequencies are omitted for brevity.

\subsubsection{Lossless model} \label{sec:ApproxLossless}
Note that in our model, we neglect intrinsic dissipation and dephasing processes, focusing on the unitary evolution of the system.
When dissipation occurs on a timescale much longer than the characteristic photon emission duration, it affects the emission dynamics, and hence the emitted photon characteristics, only negligibly.
Thus, in this scenario, the lossless model offers a valid approximation for the system state evolution, grasping its essential features.

Indeed, for the photon leakage rates to waveguides and the excitation exchange rates between the resonators and transmons of the order of several MHz, one estimates the photon emission duration to be in the sub-$\mu\mathrm{s}$ range.
On the other hand, the internal quality factors of modern lumped-element resonators exceed $10^6$~\cite{lilishi2022, crowley2023, dhundhwal2025apl, gupta2025prappl}, reaching up to $5 \times 10^7$~\cite{acharya2026arxiv}, which corresponds to photon lifetimes on the order of hundreds of microseconds.
As mentioned in Sec.~\ref{sec:Implementation}, the excited-state lifetimes of modern transmons routinely surpass 100 $\mu\mathrm{s}$, attaining a millisecond mark recently demonstrated in Ref.~\cite{bland2025nat}; see also the recent review by H. Terai \textit{et al.}~\cite{terai2026apex} and references therein.
Hence, assuming the setup components have state-of-the-art characteristics, dissipation is at least two orders of magnitude slower than emission.

\section{System state evolution}
\label{sec:Evolution}

We now proceed to study the evolution of the system quantum state.
The qubits are prepared in their excited states, while the resonator modes and waveguides are in their vacuum states.
Hence, the system initially contains two excitations.

The system Hamiltonian expressed by Eqs.~\eqref{eq:FullHamiltonian}--\eqref{eq:WaveguideHamiltonian} satisfies $[\hat H, \hat{\mathcal{N}}_s] = 0$, with $\textstyle \hat{\mathcal{N}}_s = \hat a^\dag_s \hat a_s + \hat \sigma^+_s \hat \sigma^-_s + \int \dd\omega \, \hat w^\dag_{s\omega} \hat w_{s\omega}$ being the operator of the excitation number in subsystem $s$.
This property implies that the number of excitations in each subsystem, as well as the total number of excitations in the system, is conserved.
Since dissipation in the considered system is assumed to be negligible, as discussed in Sec.~\ref{sec:Model}, its quantum state remains confined within the two-excitation domain.

Here, we also assume that the number of thermal excitations in the system is negligible.
Indeed, circuit QED setups typically operate at temperatures of about 10 mK to 30 mK.
Given that the working frequencies of circuit QED systems are a few GHz or higher, one estimates the average number of thermal excitations in the system to be less than $10^{-2}$.

\subsection{Time-dependent state vector and evolution equations}
\label{sec:Wavefunction}

Therefore, at an arbitrary instant $t$, the quantum state of the system $|\varPsi_{AB}(t)\rangle$ is expressed as a superposition of all possible two-excitation states in which the system can be found.
Hence, the state vector $|\varPsi_{AB}(t)\rangle$ reads as
\begin{widetext}
\begin{equation} \label{eq:Wavefunction}
 \begin{split}
   |\varPsi_{AB}(t)\rangle = {}& \int \dd\omega_A \int \dd \omega_B \, \varPhi(\omega_A, \omega_B; t) \hat w^\dag_{A\omega_A} \hat w^\dag_{B\omega_B}\vacket
   + \sum_{s} \int \dd\omega \, \mathcal{X}_{s\bar{s}}(\omega; t) \hat a^\dag_{s} \hat w^\dag_{\bar{s}\omega} \vacket \\
   &{}+ \sum_{s} \int \dd\omega \, \mathcal{Y}_{s\bar{s}}(\omega; t) \hat\sigma^+_{s} \hat w^\dag_{\bar{s}\omega} \vacket
   + \sum_{s} \mathcal{W}_{s\bar{s}}(t) \hat\sigma^+_s \hat a^\dag_{\bar{s}} \vacket
   + \mathcal{R}(t) \hat a^\dag_A \hat a^\dag_B \vacket + \mathcal{Q}(t) \hat\sigma^+_A \hat\sigma^+_B \vacket,
 \end{split} 
\end{equation}
\end{widetext}
where $\vacket$ denotes the vacuum state of the \emph{entire} system---the state with transmons residing in their ground states and both resonators and waveguides in their vacuum states.

The first term in Eq.~\eqref{eq:Wavefunction} corresponds to the state with both waveguides accommodating a single photon, whereas the resonators are in the vacuum state, and the qubits reside in their ground states.
Here, $\varPhi(\omega_A, \omega_B; t)$ denotes the joint spectral amplitude of the emitted photons.
Its modulus squared, $\mathcal{J}(\omega_A, \omega_B; t) = |\varPhi(\omega_A, \omega_B; t)|^2$, is often referred to as the joint spectral intensity (JSI)~\cite{zielnicki2018jmo} and represents the probability density distribution of finding at an instant $t$ photons with frequencies $\omega_A$ and $\omega_B$ in waveguides $A$ and $B$, respectively.
Hence, the probability $\varPi(t)$ that both waveguides host photons at an instant $t$ is determined as
\begin{equation} \label{eq:TwoPhotonProbability}
    \varPi(t) = \int \dd\omega_A \int \dd\omega_B \, \mathcal{J}(\omega_A, \omega_B; t). 
\end{equation}

The next two terms in Eq.~\eqref{eq:Wavefunction} describe the states with only one waveguide hosting a single photon, while the other waveguide is in the vacuum state.
The amplitude $\mathcal{X}_{s\bar{s}}(\omega; t)$ corresponds to the state with ground-state qubits and the resonator $s$ in the single-photon state. $\mathcal{Y}_{s\bar{s}}(\omega; t)$ denotes the amplitude of the state with both resonators void of photons and the excited-state qubit in the subsystem $s$.

The last three terms in Eq.~\eqref{eq:Wavefunction} describe the states with waveguides containing no photons.
 The amplitude $\mathcal{W}_{s\bar{s}}(t)$ corresponds to the state with the excited-state qubit in subsystem $s$ and the single-photon state in the resonator in subsystem $\bar{s}$, while the qubit in subsystem $\bar{s}$ resides in its ground state and the resonator in subsystem $s$ is in the vacuum state.
In the penultimate term, $\mathcal{R}(t)$ is the amplitude of the state with both resonator modes being in the single-photon state, while the qubits are in their ground states.
Finally, $\mathcal{Q}(t)$ in the last term denotes the amplitude of the state with both qubits in their excited states, whereas the resonators are in the vacuum state.

\begin{figure*}
    \centering
	\includegraphics{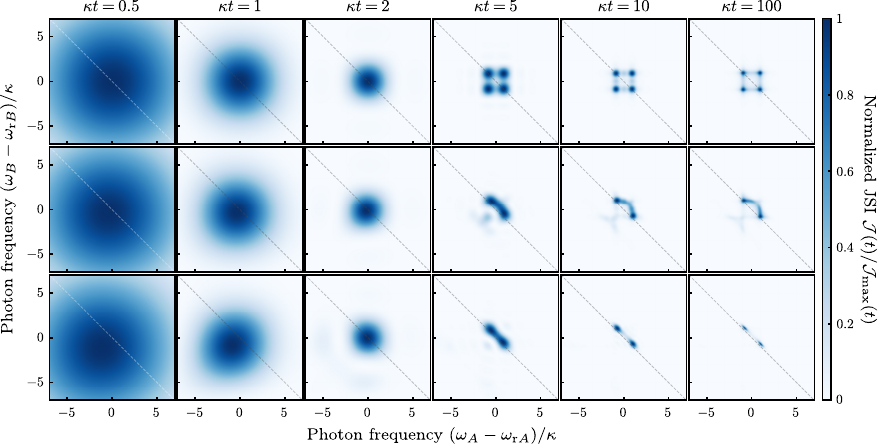}
	\caption{The snapshots of the JSI of the emitted photons for the different emission stages and various cross-Kerr coupling rates: $\chi/g = 0$ (upper row), $\chi/g = 2$ (middle row), and $\chi/g = 6$ (bottom row).
    As explained in the main text, we set the same-type couplings in both subsystems to be identical: $(g_{A} = g_{B}) = g$, $(\chi_{AB} = \chi_{BA}) = \chi$, and $(\kappa_A = \kappa_B) = \kappa$.
    For computations, we set $g/\kappa = 1$.
    We also assume the resonator-transmon interactions to be resonant, $(\Delta_A = \Delta_B) = 0$.
    To aid the visualization, each JSI plot are normalized to its maximum value $\textstyle \mathcal{J}_\mathrm{max}(t) = \max_{\omega_A, \omega_B} \mathcal{J}(\omega_A,\omega_B;t)$ for a given $t$. 
    Dashed diagonal lines correspond to perfect frequency anticorrelation $\omega_A + \omega_B = \omega_{\res A} + \omega_{\res B}$.
    \label{fig:JSIEvolution}}
\end{figure*}

At the initial moment ($t=0$), the state of the system $|\Psi_0\rangle \equiv |\varPsi_{AB}(t=0)\rangle$ reads as:
\begin{equation} \label{eq:InitialState}
  |\Psi_0\rangle = \hat \sigma^+_A \hat \sigma^+_B \vacket,
\end{equation}
which corresponds to the state with both qubits in their excited states, while the resonators and the waveguides accommodate no photons.

The evolution of the probability amplitudes in Eq.~\eqref{eq:Wavefunction} is governed by the set of equations of motion as follows:
\begin{widetext}
\begin{subequations} \label{eq:EvolutionEquations}
\begin{equation} \label{eq:EvolutionEqPhi}
    \left[\ii \pdt - (\omega_{A} + \omega_{B})\right] \varPhi(\omega_A,\omega_B; t) = f_{A} \, \mathcal{X}_{AB}(\omega_B; t) + f_{B} \, \mathcal{X}_{BA}(\omega_A; t),
\end{equation}
\begin{equation} \label{eq:EvolutionEqY}
    \left[\ii \pdt - (\omega + \omega_{\qb s})\right] \mathcal{Y}_{s\bar{s}}(\omega; t) = f_{\bar{s}} \mathcal{W}_{s\bar{s}}(t) + g_{s} \mathcal{X}_{s\bar{s}}(\omega; t), 
\end{equation}
\begin{equation} \label{eq:EvolutionEqX}
    \left[\ii\pdt - (\omega + \omega_{\res s}) + \ii\frac{\kappa_s}{2}\right] \mathcal{X}_{s\bar{s}}(\omega_s; t) = f_{\bar{s}} \mathcal{R}(t) + g_s \mathcal{Y}_{s\bar{s}}(\omega; t),
\end{equation}
\begin{equation} \label{eq:EvolutionEqW}
 \left[\ii \pdt - (\omega_{\qb s} + \omega_{\res \bar{s}} - \chi_{s\bar{s}}) + \ii \frac{\kappa_{\bar{s}}}{2}\right] \mathcal{W}_{s\bar{s}}(t) = g_s \mathcal{R}(t) + g_{\bar{s}} \mathcal{Q}(t),
\end{equation}
\begin{equation} \label{eq:EvolutionEqR}
 \left[\ii\pdt - (\omega_{\res A} + \omega_{\res B}) + \ii\frac{\kappa_A + \kappa_B}{2}\right] \mathcal{R}(t) = g_A \mathcal{W}_{AB}(t) + g_B \mathcal{W}_{BA}(t),
\end{equation}
\begin{equation} \label{eq:EvolutionEqQ}
 \left[\ii\pdt - (\omega_{\qb A} + \omega_{\qb B}) \right]\mathcal{Q}(t) = g_{A} \mathcal{W}_{BA}(t) + g_{B} \mathcal{W}_{AB}(t).
\end{equation}
\end{subequations}
The initial conditions for all probability amplitudes are zero, except for $\mathcal{Q}$, which is $\mathcal{Q}(t=0) = 1$.
The above set of equations of motion is derived using the approach as in Refs.~\cite{stolyarov2019, stolyarov2022}.
\end{widetext}

\subsection{Emission dynamics}
\label{sec:EmissionDynamics}

\begin{figure} 
    \centering
    \includegraphics{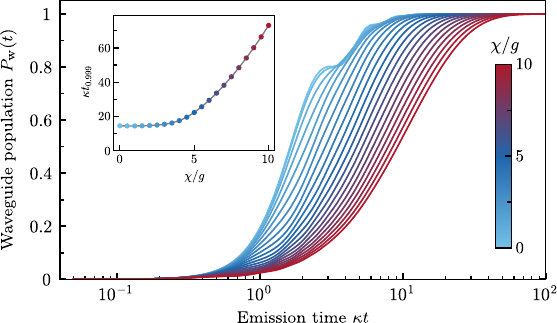}
    \caption{Effect of the cross-Kerr coupling on the waveguide population dynamics.
    Dependence of the waveguide population $P_{\wg}(t) = P_{\wg A}(t) = P_{\wg B}(t)$ on the dimensionless emission time $\kappa t$ for different values of the dimensionless cross-Kerr coupling rate $\chi/g$ encoded as a color gradient.
    As in Fig.~\ref{fig:JSIEvolution}, we consider the resonant regime of transmon--resonator couplings, $(\Delta_A = \Delta_B) = 0$, and identical rates of the same-type couplings, $(g_A = g_B) = g$, $(\chi_{AB} = \chi_{BA}) = \chi$, and $(\kappa_A = \kappa_B) = \kappa$, with $g/\kappa = 1$.
    Inset shows the dependence of the dimensionless time $\kappa t_{0.999}$ at which waveguide population $P_{\mathrm w}$ reaches $0.999$ on the ratio $\chi/g$.
    }
    \label{fig:WaveguidePopulation}
\end{figure}

By numerically solving (see details in Appendix~\ref{app:NumericalMethods}) the set of evolution equations~\eqref{eq:EvolutionEquations}, we determine the quantum state of the system at an arbitrary instant and obtain a real-time picture of the emission dynamics.
Figure~\ref{fig:JSIEvolution} shows the set of snapshots of the JSIs at specific instants for the different cross-Kerr coupling rates.
The animations of the JSIs are available in the Supplemental Material~\cite{supplementary}.
For definiteness and to simplify the analysis, here and throughout the rest of the manuscript, we consider the case of identical rates of the same-type couplings in both subsystems, i.e., $(g_{A} = g_{B}) = g$, $(\chi_{AB} = \chi_{BA}) = \chi$, and $(\kappa_A = \kappa_B) = \kappa$. 

Figure~\ref{fig:JSIEvolution} reveals that switching on cross-Kerr coupling between the subsystems modifies the spectral characteristics of the emitted photons.
It leads to the onset of frequency anticorrelation of the emitted photons, manifesting itself in the JSI accumulation along the line determined by the condition $\omega_A + \omega_B = \omega_{\res A} + \omega_{\res B}$.
We will characterize this effect quantitatively in Sec.~\ref{sec:FrequencyCorrelation}.

Once the time-dependent state vector $|\varPsi_{AB}(t)\rangle$ is determined, one is able to compute the waveguide populations at an arbitrary instant of time.
Using the probability amplitudes in Eq.~\eqref{eq:Wavefunction}, we express the population $P_{\wg s}(t)$ of the waveguide in the subsystem $s$ at instant $t$ as
\begin{equation} \label{eq:WaveguidePopulation}
   \begin{split}
     P_{\wg s}(t) = & \, \int \dd\omega \langle \varPsi_{AB}(t)|\hat w^\dag_{s\omega} \hat w_{s\omega}|\varPsi_{AB}(t)\rangle \\ 
     = & \, \varPi(t) + \int \dd\omega |\mathcal{X}_{\bar{s}s}(\omega;t)|^2 + \int \dd\omega |\mathcal{Y}_{\bar{s}s}(\omega;t)|^2,
   \end{split}
\end{equation}
where $\varPi(t)$ is given by Eq.~\eqref{eq:TwoPhotonProbability}.

Figure~\ref{fig:WaveguidePopulation} demonstrates the effect of the cross-Kerr coupling between the subsystems on the waveguide population dynamics.
In the regime of identical same-type couplings we consider, the waveguide populations are equal, $P_{\wg A}(t) = P_{\wg B}(t) = P_{\wg}(t)$. 
The computations reveal that the growth of the waveguide population $P_{\wg}(t)$ with time becomes slower as the cross-Kerr coupling rate increases; c.f. inset in Fig.~\ref{fig:WaveguidePopulation}.
This effect can be explained as follows.
The cross-Kerr coupling detunes the resonators from the transmons, which in turn suppresses the exchange of excitations between them in both subsystems.
This suppression of the excitation exchange between the transmons and resonators slows down the emission of photons into the waveguides.

\subsection{Photon cross-correlations}
\label{sec:CrossCorrelations}

\begin{figure}[t!] 
	\centering
	\includegraphics{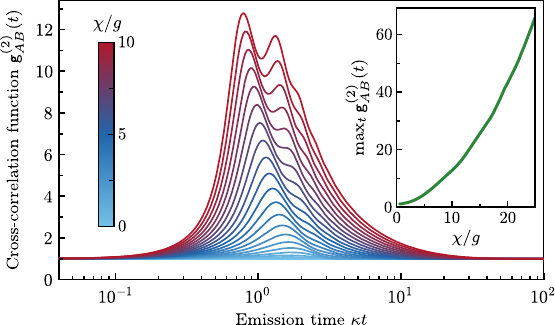}
	\caption{Cross-correlation function $\mathtt{g}^{(2)}_{AB}(t)$ versus the dimensionless emission time $\kappa t$ for different values of the (dimensionless) cross-Kerr coupling rate $\chi/g$ encoded as a color gradient.
	Inset shows the dependence of the cross-correlation function maximum $\textstyle \max_t \mathtt{g}^{(2)}_{AB}(t)$ on the ratio $\chi/g$.
	The system parameters used for calculations are the same as in Fig.~\ref{fig:WaveguidePopulation}.
	\label{fig:CrossCorrelations}}
\end{figure}

As discussed in Sec.~\ref{sec:Principle}, the cross-Kerr interaction couples the emission dynamics of the two subsystems.
To characterize the resulting correlations of the emitted photons, we consider the normalized equal-time photon-number cross-correlation function
\begin{equation} \label{eq:CrossCorrFunc}
	\mathtt{g}^{(2)}_{AB}(t) = \frac{\langle \varPsi_{AB}(t)|\hat n_{\wg A} \hat n_{\wg B}|\varPsi_{AB}(t)\rangle}{\langle \varPsi_{AB}(t)|\hat n_{\wg A}|\varPsi_{AB}(t)\rangle\langle \varPsi_{AB}(t)|\hat n_{\wg B}|\varPsi_{AB}(t)\rangle},
\end{equation}
where $\textstyle \hat n_{\wg s} = \int \dd\omega \, \hat w^\dag_{s\omega} \hat w_{s\omega}$ is the total photon-number operator of waveguide $s$.
Since each subsystem initially contains one excitation and its excitation number is conserved, each waveguide can accommodate at most one photon.

Equation~\eqref{eq:TwoPhotonProbability} suggests that $\textstyle \langle\varPsi_{AB}(t)|\hat n_{\wg A} \hat n_{\wg B}|\varPsi_{AB}(t)\rangle = \varPi(t)$.
By plugging this result, along with the first line in Eq.~\eqref{eq:WaveguidePopulation}, into Eq.~\eqref{eq:CrossCorrFunc}, we obtain
\begin{equation} \label{eq:CrossCorrFunc2}	
  \mathtt{g}^{(2)}_{AB}(t)= \frac{\varPi(t)}{P_{\wg A}(t) P_{\wg B}(t)}.
\end{equation}
The expression above compares the joint probability of photon occupation in both waveguides with the product of the corresponding marginal probabilities.
In the outgoing-waveguide model considered here, these probabilities describe emission accumulated up to time $t$.
Thus, $\mathtt{g}^{(2)}_{AB}(t)>1$ indicates that both photons have been emitted by time $t$ with a probability exceeding that expected for statistically independent emission events with the same marginal probabilities.
Equivalently, the presence of a photon in one waveguide increases the conditional probability that the other waveguide is also occupied.
This observable characterizes correlations between the accumulated waveguide populations.

When the subsystems $A$ and $B$ are uncoupled, i.e., there are no cross-Kerr interactions between them, $(\chi_{AB} = \chi_{BA}) = 0$, they evolve independently.
In this regime, the state vector of the system can be represented as a product $|\varPsi_{AB}(t)\rangle = |\psi_A(t)\rangle \otimes |\psi_B(t)\rangle$, where $|\psi_s(t)\rangle$ denotes the state of the subsystem $s$ at an instant $t$.
Consequently, $\varPi(t) = P_{\wg A}(t)P_{\wg B}(t)$, yielding $g^{(2)}_{AB}(t) = 1$.
Switching on the cross-Kerr interaction produces departure from this independent-emission behavior.
Figure~\ref{fig:CrossCorrelations} shows an enhancement of the joint occupation probability during emission, with the maximum of $g^{(2)}_{AB}(t)$ increasing with the cross-Kerr coupling strength over the parameter range considered.

At long times, both excitations have been transferred to the waveguides, so that $P_{\wg A}(t)$, $P_{\wg B}(t)$, and $\varPi(t)$ approach unity. Hence, $\mathtt{g}^{(2)}_{AB}(t)$ also approaches unity.
This limit reflects the definite final photon number in each output channel and does not imply that the emitted photons are uncorrelated in frequency or emission time.
The spectral correlations and entanglement of the output state are examined in Sec.~\ref{sec:OutputState}.

\section{Output state of photons}
\label{sec:OutputState}

The considered system ultimately reaches the state $|\Psi_\mathrm{out}\rangle$ when the transmon--resonator dimer contains no excitations, as it has completely emitted the photons into the output waveguides, where they propagate as free excitations. 
The output state $|\Psi_\mathrm{out}\rangle$ reads as
\begin{widetext}
\begin{equation}
    |\Psi_\mathrm{out}\rangle = \int \dd\omega_A \int \dd\omega_B \, \ee^{-\ii (\omega_A + \omega_B) t} \Phi_\mathrm{out}(\omega_A, \omega_B) \hat w^\dag_{A\omega_A} \hat w^\dag_{B\omega_B} \vacket, 
\end{equation}
where $\textstyle \Phi_\mathrm{out}(\omega_A,\omega_B) = \lim_{t \rightarrow \infty} \ee^{\ii (\omega_A + \omega_B) t} \varPhi(\omega_A, \omega_B; t)$ denotes the JSA of the output photon pair.
\end{widetext}

For the pure output biphoton state we consider, if its JSA is factorizable as $\Phi_\mathrm{out}(\omega_A,\omega_B) = u(\omega_A) v(\omega_B)$, there is no frequency entanglement between the emitted photons; otherwise, the photons are entangled.

\subsection{Frequency correlation}
\label{sec:FrequencyCorrelation}

\begin{figure} 
    \centering
    \includegraphics{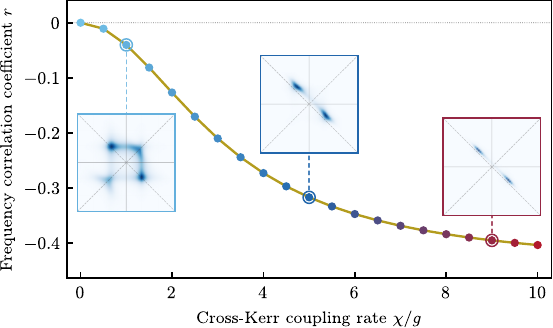}
    \caption{Dependence of the dimensionless frequency correlation coefficient $r$ on the dimensionless cross-Kerr coupling rate $\chi/g$.
    Insets show the JSIs of the output photons for specific values of $\chi/g$.
    For computations, we use the same parameters as in Fig.~\ref{fig:JSIEvolution}.
    }
    \label{fig:FrequencyCorrelation}
\end{figure}

Here, we aim to demonstrate the onset of frequency correlation of the emitted photons quantitatively.
To characterize frequency correlations between the emitted photons, we use the Pearson frequency-correlation coefficient
\begin{equation} \label{eq:FrequencyCorrelation}
    r = \frac{\mathrm{Cov}(\omega_A, \omega_B)}{\sqrt{\mathrm{Var}(\omega_A) \mathrm{Var}(\omega_B)}},
\end{equation}
where $\mathrm{Var}(\omega_s) = \int \dd \omega_A \int \dd\omega_B (\omega_s - \bar\omega_s)^2 J_\mathrm{out}(\omega_A, \omega_B)$ and $\mathrm{Cov}(\omega_A, \omega_B) = \int \dd\omega_A \int \dd\omega_B \, (\omega_A - \bar\omega_A) (\omega_B - \bar\omega_B) J_\mathrm{out}(\omega_A, \omega_B)$.
Here, $J_\mathrm{out}(\omega_A, \omega_B) = |\Phi_\mathrm{out}(\omega_A, \omega_B)|^2$ is the JSI of the emitted photons, and $\textstyle \bar\omega_s = \int \dd\omega_A \int \dd\omega_B \, \omega_s J_\mathrm{out}(\omega_A, \omega_B)$ is the average frequency of the output photon in the waveguide $s$.
Positive and negative values of $r$ indicate positive and negative linear frequency correlations, respectively.
The magnitude $|r|$ characterizes the strength of these linear correlations, with $r = -1$ corresponding to perfect linear anticorrelation.
A vanishing value indicates the absence of linear correlation but does not, in general, imply statistical independence of the photon frequencies.

The results of the computations clearly demonstrate the defining effect of the cross-Kerr interaction on the JSI properties.
In agreement with the qualitative considerations in Sec.~\ref{sec:Principle}, without cross-Kerr coupling, the two subsystems evolve independently, and the output joint spectral intensity factorizes, yielding $r=0$.
Switching on the cross-Kerr interaction produces frequency anticorrelation, as indicated by the negative values of $r$ in Fig.~\ref{fig:FrequencyCorrelation}.
Over the parameter range considered, increasing the cross-Kerr coupling strength increases the magnitude of $r$, indicating stronger frequency anticorrelation.

Note that, in general, the frequency-correlation coefficient $r$ is not an entanglement measure.
It characterizes linear correlations in the JSI of the emitted photons, however, it is insensitive to entanglement encoded in a nonseparable joint phase.
This limitation of intensity-based correlation characterization was demonstrated for transverse spatial entanglement in~\cite{chan2007pra} and spectral entanglement in~\cite{jeronimo-moreno2009pra}.
Bell \textit{et al.}~\cite{bell2020} further investigated spectral-phase correlations that are inaccessible for JSI measurements and demonstrated their detection using frequency-resolved multiphoton emission.
Hence, although a nonzero $r$ is sufficient to indicate frequency entanglement for the \emph{pure} biphoton states considered here, $r = 0$ does not imply separability.
Moreover, for \emph{mixed} states, $r \neq 0$ does not necessarily imply entanglement. 
A mixed state may exhibit frequency correlations while still being separable.
We therefore turn to the Schmidt decomposition of the JSA to characterize the spectral entanglement of the emitted photons.
We provide this analysis below.

\subsection{Schmidt decomposition}
\label{sec:Schmidt}

Nonfactorizability of the JSA $\Phi_\mathrm{out}(\omega_A,\omega_B)$ and hence frequency entanglement of the emitted photons can be determined via the Schmidt decomposition~\cite{law2000}:
\begin{equation} \label{eq:SchmidtDecomposition}
    \Phi_\mathrm{out}(\omega_A,\omega_B) = \sum_{k\geq 1} \sqrt{\lambda_k} \, u_k(\omega_A) v_k(\omega_B).
\end{equation}
The decomposition coefficients are nonnegative $\lambda_k \geq 0$ and satisfy the condition $\textstyle \sum_{k} \lambda_k = 1$.
In our analysis, we assume that the Schmidt coefficients are ordered in descending magnitude, $\lambda_{k} \geq \lambda_{k+1}$.
If there is only one nonzero Schmidt coefficient, i.e., $\lambda_1 = 1$ and  $\lambda_{k\geq 2} = 0$, the output JSA is factorizable, and the emitted photons are not frequency-entangled.
Two or more nonzero coefficients indicate frequency entanglement.

\begin{figure*}[t!] 
	\centering
	\includegraphics{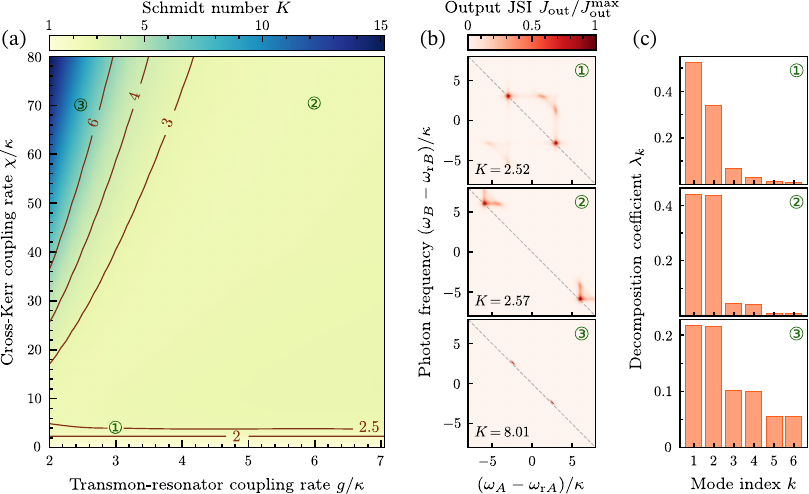}
	\caption{
    (a) Dependence of the Schmidt number $K$ of the emitted photon pair on the (dimensionless) rates of cross-Kerr $\chi/\kappa$ and transmon--resonator couplings $g/\kappa$.
    Markers: \ding{172} is a point with $K=2.52$ at $g/\kappa=3$, $\chi/\kappa=4$, whose Schmidt spectrum decays monotonically; 
    \ding{173} denotes a point at $g/\kappa=6$ and $\chi/\kappa=70$ with nearly the same Schmidt number ($K=2.57$) as \ding{172} but exhibiting a completely different Schmidt spectrum, where two coefficients dominate;
    \ding{174} marks a point with a large Schmidt number ($K=8.01$) at $g/\kappa=2.5$ and $\chi/\kappa=70$.
    For each marked point, we plot 
    (b) the corresponding JSI normalized to its peak value $\textstyle J^\mathrm{max}_\mathrm{out} = \max_{\omega_A, \omega_B} J_\mathrm{out}(\omega_A, \omega_B)$ and (c) the first six Schmidt coefficients $\lambda_k$.
    \label{fig:SchmidtNumberDependence}}
\end{figure*}

The Schmidt modes form a complete set of orthonormal functions.
They are determined by solving the eigenvalue problem~\cite{law2000}:
 \begin{equation} \label{eq:EigenvalueProblem}
   \int \dd\nu \, \varUpsilon_u(\nu,\omega) u_k(\nu) = \lambda_k u_k(\omega), \quad
   \int \dd\nu \, \varUpsilon_v(\nu,\omega) v_k(\nu) = \lambda_k v_k(\omega).
 \end{equation}
The integral kernel $\varUpsilon_{u, v}(\nu,\omega)$ is defined as
 \begin{equation} \label{eq:IntegralKernels}
    \varUpsilon_{u}(\nu,\omega) = \int\dd\varpi \, \Phi^*_\mathrm{out}(\nu,\varpi) \Phi_\mathrm{out}(\omega,\varpi), \quad
    \varUpsilon_{v}(\nu,\omega) = \int\dd\varpi \, \Phi^*_\mathrm{out}(\varpi,\nu) \Phi_\mathrm{out}(\varpi,\omega).
 \end{equation}

In practice, the decomposition~\eqref{eq:SchmidtDecomposition} is obtained not by solving the integral eigenvalue problem~\eqref{eq:EigenvalueProblem} directly, but by singular value decomposition (SVD) of the discretized JSA $\Phi_\mathrm{out}(\omega_A,\omega_B)$ of the emitted photon pair; see details in Appendix~\ref{app:SVDDetails}.
Both approaches are equivalent~\cite{ekert1995, zielnicki2018jmo} and underlie the numerical Schmidt analyses of parametric down-conversion~\cite{law2004, rohde2007, brecht2015}.

As a figure of merit characterizing the strength of entanglement of the emitted photons, we use the Schmidt number
\begin{equation} \label{eq:SchmidtNumber}
	K = \Bigl(\sum_{k} \lambda_k^2\Bigr)^{-1},
\end{equation}
which measures the effective number of populated Schmidt modes~\cite{law2000, law2004}.
When $K=1$, the output photons are not entangled, while $K>1$ indicates frequency entanglement.

Figure~\ref{fig:SchmidtNumberDependence}(a) demonstrates the dependence of the Schmidt number of the output photon state on the strength of cross-Kerr coupling $\chi$ between the subsystems and the transmon--resonator coupling rate $g$ in the resonant regime $\Delta_A = \Delta_B = 0$.
The output photons become frequency-entangled once the cross-Kerr coupling is switched on.
As the cross-Kerr coupling rate $\chi$ grows, the Schmidt number $K$ rises steeply and surpasses the value $K=2$ at $\chi/\kappa\approx 2.3$, almost independently of the transmon--resonator coupling rate $g$.

The examples in Figs.~\ref{fig:SchmidtNumberDependence}(b) and \ref{fig:SchmidtNumberDependence}(c) demonstrate that varying the transmon–resonator and cross-Kerr coupling strengths modifies both the JSI and the distribution of weight among the Schmidt modes.
Hence, the considered source enables access to biphoton states with qualitatively different spectral and entanglement structures.
At point \ding{174}, the Schmidt weight is distributed over many modes, yielding a comparatively large effective dimensionality.
In contrast, points \ding{172} and \ding{173} have comparable Schmidt numbers, $K \approx 2.52$ and $K \approx 2.57$.
However, their Schmidt spectra differ significantly: at point \ding{172}, the decomposition coefficients decrease gradually, whereas at point \ding{173}, two nearly equal coefficients dominate over the remaining contributions.
This result highlights that the Schmidt number alone does not effectively distinguish between these different distributions of entanglement.

Notably, the regime represented by point \ding{173} is particularly relevant to two-dimensional spectral encoding.
An ideal maximally entangled state in two pairs of orthonormal modes has precisely two nonzero Schmidt coefficients, $\lambda_1 = \lambda_2 = 1/2$.
Therefore, a spectrum dominated by two nearly equal coefficients indicates proximity to this structure.
Together with the well-separated spectral features in Fig.~\ref{fig:SchmidtNumberDependence}(b), this behavior motivates examining whether the emitted biphoton state approximates the frequency-bin Bell state.
In the following section, we quantify the overlap of the emitted state with a two-dimensional maximally entangled state and investigate how adjusting the coupling strengths and transmon–resonator detuning can improve this overlap.

\section{Frequency-bin entangled biphotons} \label{sec:FrequencyBinEntanglement}

The analysis in Sec.~\ref{sec:Schmidt} demonstrates that the proposed source generates biphotons with qualitatively different Schmidt spectra.
In particular, the state corresponding to point~\ding{173} in Fig.~\ref{fig:SchmidtNumberDependence} is
dominated by two nearly equally weighted Schmidt contributions,
whereas other parameter choices distribute the entanglement over
a larger number of modes.
This distinction is essential for two-dimensional spectral encoding.
In this encoding, two orthogonal single-photon spectral modes serve as the logical basis states of each qubit~\cite{lu2023opt}.
When these modes occupy well-separated frequency intervals, they define a frequency-bin qubit whose information is carried by the relative amplitudes and the phase of their superposition.
For a photon pair, with one photon in each output waveguide, a natural target is a Bell state---a maximally entangled state of the two frequency-encoded qubits.
In the frequency-anticorrelated case, this corresponds to an equal-weight superposition of two configurations: one photon occupying the lower-frequency mode and the other the higher-frequency mode, or vice versa.

Such states provide a resource for distributing entanglement between spatially separated receivers~\cite{khodadadkashi2025lsa} and for conditional
preparation of photonic spectral states.
Remote preparation through coherent mode-selective measurement of one photon in an entangled pair has been demonstrated for temporal modes~\cite{ansari2020optexpr}.
Applied to a frequency-bin Bell pair, the same principle allows a suitable measurement of one photon to conditionally prepare its partner in a prescribed superposition of the two frequency bins.
These prospects motivate assessing how closely the emitted biphoton approximates an ideal Bell pair.
First, we quantify its proximity to a maximally entangled state within a two-dimensional local spectral subspace.
Then, we determine whether the corresponding modes can be identified with separated frequency bins.

\subsection{Bell-state fidelity}

To define the target Bell states, we introduce two sets of orthonormal
single-photon spectral modes in each waveguide.
In waveguide~$A$, these states are $|\xi_j\rangle_A = \int \dd\omega\, \xi_j(\omega) \hat w^\dagger_{A\omega}|0\rangle_A$, where $j \in \{1,2\}$ is a mode index.
Similarly, in waveguide~$B$, one has $|\eta_j\rangle_B = \int \dd\omega\, \eta_j(\omega) \hat w^\dagger_{B\omega}|0\rangle_B$.
The spectral amplitudes $\xi_j(\omega)$ and $\eta_j(\omega)$ satisfy the orthonormality condition $\int \dd\omega \, \xi_i^*(\omega)\xi_j(\omega) = \int \dd\omega \,
\eta_i^*(\omega)\eta_j(\omega) =\delta_{ij}$.
Each pair of modes defines a two-dimensional single-photon subspace in its respective waveguide.
For any such choice of modes, an ideal Bell state can be represented as
\begin{equation} \label{eq:BellTarget}
	|\BellState_{\xi,\eta,\varphi}\rangle = \frac{|\xi_1\rangle_A|\eta_1\rangle_B + e^{\ii \varphi}|\xi_2\rangle_A|\eta_2\rangle_B}{\sqrt{2}},
\end{equation}
where $\varphi$ denotes the relative phase between the two product-state contributions.
Their equal amplitudes and local orthogonality make this state maximally entangled within the selected subspaces.

The concentration of Schmidt weight into two nearly equally populated mode pairs suggests proximity of the emitted biphoton state to a state of the form of Eq.~\eqref{eq:BellTarget}.
However, the Schmidt number alone does not quantify this proximity.
To make the comparison quantitative, we employ the fidelity of the emitted state with the best-matching ideal Bell state encoded in two spectral modes per photon:
\begin{equation}
\label{eq:BellFidelity}
\begin{split}
	F = & \, \max_{\bm{\beta}}\left|\langle \BellState_{\xi,\eta,\varphi}|\Psi_{\mathrm{out}}\rangle\right|^2 \\
	= & \, \frac{1}{2} \left(\sqrt{\lambda_1}+\sqrt{\lambda_2}\right)^2,	
\end{split}
\end{equation}
where the maximum is found in the entire set of Bell-state parameters $\bm{\beta} = \{\xi_j, \eta_j, \varphi\}$.
It is a two-dimensional version of the fully entangled fraction~\cite{bennett1996mixed, horodecki1999, vidal2000, grondalski2002}.

For the pure output state with Schmidt decomposition given in Eq.~\eqref{eq:SchmidtDecomposition}, the maximum is attained by choosing $\xi_j = u_j$ and $ \eta_j = v_j$ for $j \in \{1,2\}$, corresponding to the two largest Schmidt coefficients.
Since the Schmidt amplitudes $\sqrt{\lambda_j}$ are real and nonnegative, this choice gives the optimal relative phase $\varphi=0$.
Hence, the corresponding best-matching Bell state is $|\BellState_*\rangle = (|u_1\rangle_A|v_1\rangle_B + |u_2\rangle_A|v_2\rangle_B)/\sqrt{2}$.
The overlap of this state with the emitted state $|\Psi_\mathrm{out}\rangle$ is determined by the second line in Eq.~\eqref{eq:BellFidelity}.

Equation~\eqref{eq:BellFidelity} evaluates the overlap with the full output state.
Therefore, it retains the penalty associated with population in additional Schmidt modes and involves no postselection onto a two-mode sector.
Attaining the ideal fidelity $F = 1$ requires $\lambda_1 = \lambda_2 = 1/2$, with all remaining Schmidt coefficients equal to zero.
To distinguish the mechanisms that reduce the fidelity, we introduce the weight $p_2 = \lambda_1 + \lambda_2$ of the dominant two-mode sector.
Projection onto the corresponding local Schmidt subspaces\footnote{Restricting a multimode entangled state to a two-dimensional sector, renormalizing, and assessing its Bell quality is a standard approach, which was applied to orbital-angular-momentum~\cite{zhang2017ncom, bernecker2025pra}, energy-time~\cite{steinlechner2017ncom}, and frequency-bin~\cite{chen2020pra} encodings. The underlying projection onto a selected Schmidt sector goes back to Ref.~\cite{bennett1996concentrating}.} produces, with a probability $p_2$, the normalized state $|\psi_{12}\rangle = (\sqrt{\lambda_1}|u_1\rangle_A|v_1\rangle_B + \sqrt{\lambda_2}|u_2\rangle_A|v_2\rangle_B)/\sqrt{p_2}$,
whose concurrence~\cite{wootters1998} is $C_{12} = 2\sqrt{\lambda_1\lambda_2}/(\lambda_1+\lambda_2)$.
Thus, fidelity can be expressed as a product of the probability of occupying the two-mode sector and the Bell-state fidelity of the output state projected onto that sector:
\begin{equation} \label{eq:BellFidelityFactorization}
	F = p_2 \, \frac{1 + C_{12}}{2}.
\end{equation}
Equivalently, the infidelity $1-F$ is expressed as
\begin{equation} \label{eq:Infidelity}
 \begin{aligned}
	1-F	& = (1-p_2)+\frac{p_2}{2}(1-C_{12})
	\\
	& = \sum_{k\geq 3}\lambda_k + \frac{1}{2} \left(\sqrt{\lambda_1}-\sqrt{\lambda_2}\right)^2.
 \end{aligned}
\end{equation}
The first contribution is the weight in additional Schmidt modes, while the second quantifies the imbalance between the two leading contributions.
A high concurrence within the selected sector is therefore insufficient to establish a high Bell-state fidelity of the emitted state.

\begin{figure*}[t!] 
	\centering
	\includegraphics{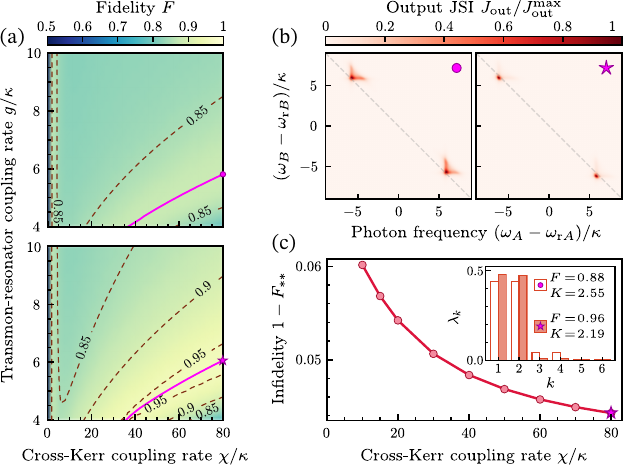}
	\caption{
		(a) Dependence of the fully entangled fraction (unrestricted Bell-state fidelity) $\Fent$ of the emitted photon pair on the rates of cross-Kerr and transmon--resonator couplings normalized on the photon leakage rate $\kappa$ at two fixed transmon--resonator detunings: $\Delta = 0$ (upper plot) and $\Delta/\kappa = -0.38$ (lower plot).
		On both plots, the solid magenta line marks the ridge---the value of the transmon--resonator coupling rate $\textstyle g_* = \mathrm{arg \, max}_g \Fent(\chi, g, \Delta)$ that maximizes the Bell-state fidelity at given $\chi$ and $\Delta$, $\textstyle \max_g F(\chi, g, \Delta) = F(\chi, g_*, \Delta)$.
		Markers denote $g_*/\kappa$ for $\chi/\kappa = 80$ and the respective $\Delta$: {$\bullet$} corresponds to $g_*/\kappa=5.83$ and $\Fent_* = 0.882$ for $\Delta = 0$, {$\star$} corresponds to $g_*/\kappa=6.08$ and $\Fent_* = 0.956$ for $\Delta/\kappa = - 0.38$, where $\textstyle F_* = F(\chi, g_*, \Delta)$.
		(b) Normalized output JSIs for the setup parameters corresponding to the marked points in plots of panel (a).
		Each JSI is normalized to its peak value $\textstyle J^\mathrm{max}_\mathrm{out} = \max_{\omega_A, \omega_B} J_\mathrm{out}(\omega_A, \omega_B)$, which is equal to $0.49\,\kappa^{-2}$ and $1.11\,\kappa^{-2}$ for {$\bullet$} and $\star$, respectively.
		(c) Dependence of infidelity $1-\Fent_{**}$ on the dimensionless cross-Kerr coupling rate $\chi/\kappa$.
		Inset shows the first six Schmidt coefficients, the fidelity $F$, and the Schmidt number $K$ for the two operating points shown in panel (b).
		\label{fig:BellFidelity}
	}
\end{figure*}

Figure~\ref{fig:BellFidelity}(a) shows how the Bell-state fidelity depends on the dimensionless cross-Kerr and transmon--resonator coupling rates $\chi/\kappa$ and $g/\kappa$ for two fixed values of the transmon--resonator detuning, $\Delta_A = \Delta_B = \Delta$.
These maps reveal a region in which suitable adjustment of the coupling rates concentrates most of the Schmidt spectrum into the first two coefficients.
Notably, this optimization differs considerably from maximizing the Schmidt number $K$.
As follows from Figs.~\ref{fig:SchmidtNumberDependence}(a) and \ref{fig:BellFidelity}(a), parameter choices that generate strongly multimode entanglement are less favorable for approximating a Bell state.
In both plots in Fig.~\ref{fig:BellFidelity}(a), the solid line marks a ridge---the transmon--resonator coupling rate $g_*(\chi,\Delta) = \operatorname*{arg\,max}_{g} F(\chi,g,\Delta)$ that maximizes the Bell-state fidelity at given $\chi$ and $\Delta$.
We denote the corresponding fidelity by $F_*(\chi,\Delta) \equiv F(\chi,g_*,\Delta)$.
Maximizing also over the transmon--resonator detuning yields the optimized fidelity $F_{**}(\chi) \equiv \max_{g,\Delta}F(\chi,g,\Delta)$, attained at $\Delta_*(\chi)$.
The dependence of the optimized fidelity $F_{**}$ on $\chi/\kappa$ is shown in Fig.~\ref{fig:BellFidelity}(c).
The optimization procedure is discussed in detail in Appendix~\ref{app:FidelityOptimization}.

In the resonant case, $\Delta=0$, the maximal fidelity of $F_* \simeq 0.882$ for $\chi/\kappa = 80$ occurs at $g_*/\kappa\simeq 5.83$.
Interestingly, introducing a small detuning between the transmon and resonator frequencies can improve the fidelity.
At $(\Delta/\kappa = \Delta_*/\kappa) \simeq -0.38$ and the same cross-Kerr coupling rate, the optimal transmon--resonator coupling rate shifts to $g_*/\kappa \simeq 6.08$, and the fidelity reaches $F_* \simeq 0.956$, resulting in more than a twofold reduction of the corresponding infidelity $1-F_*$ from approximately $0.118$ to $0.044$.

The Schmidt spectra in the inset of Fig.~\ref{fig:BellFidelity}(c) illustrate the redistribution of spectral weight underlying the fidelity improvement.
Both configurations produce two dominant, nearly equally weighted contributions, but the detuned configuration places more weight in this
leading sector.
According to Eq.~\eqref{eq:BellFidelityFactorization}, for nearly balanced coefficients with $C_{12}\simeq 1$, one has $F\simeq p_2$.
This implies that the infidelity mainly stems from nonzero populations in additional Schmidt modes.
Consistently, the Schmidt number decreases from approximately $2.55$ to $2.19$.
This comparison supports the distinction established in Sec.~\ref{sec:OutputState}\, B that the distribution of Schmidt weights, rather than just their effective number, determines how close the emitted state is to a maximally entangled two-dimensional state.

\subsection{Frequency-bin Bell-state fidelity}
\label{sec:FrequencyBinFidelity}

The JSIs demonstrated in Fig.~\ref{fig:BellFidelity}(b) exhibit two dominant, well-separated features associated with opposite frequency offsets in the output channels.
Together with the nearly balanced two-mode structure of Schmidt coefficients, these spectral features hint at a connection to frequency-bin Bell states~\cite{lu2023opt}.

Establishing a frequency-bin encoding requires identifying suitable localized spectral modes within the relevant local subspaces.
However, the unconstrained fidelity $F$ in Eq.~\eqref{eq:BellFidelity} is insufficient for this identification, as it characterizes proximity to an ideal Bell state in spectral modes that may not occupy distinct frequency intervals.
To address this issue, we complement this measure with a fidelity that explicitly incorporates the frequency-bin constraint.

\begin{figure}[t!] 
	\centering
	\includegraphics{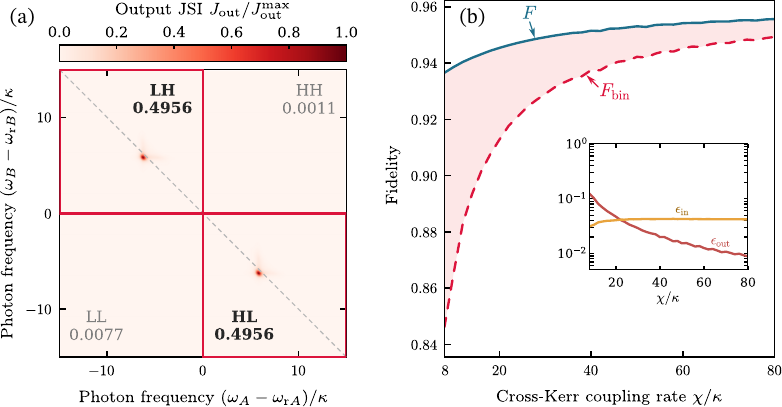}
	\caption{Identification with separated frequency bins.
		(a) Output JSI for $\chi/\kappa=80$, $g/\kappa=6.08$, and
		$\Delta/\kappa = -0.38$, which corresponds to the point $\star$ in Fig.~\ref{fig:BellFidelity}(b), divided by the bin boundary (determined by $\omega_A = \omega_{\res A}$ and $\omega_B = \omega_{\res B}$) into the four rectangles $LL$, $LH$, $HL$, $HH$ with their weights.
        The two cross-bin rectangles determined by Eq.~\eqref{eq:BinBlocks} are outlined.
        Their total weight equals to $P_\mathrm{cross}=0.991$.
		(b) Unrestricted fidelity $\Fent$ (solid line) and bin-constrained fidelity
		$\Fent_\mathrm{bin}$ (dashed line) along the ridge $g_*(\chi)$ in Fig.~\ref{fig:BellFidelity}.
        The shaded area highlights the gap between their values.
		Inset shows the two nonvanishing error terms $\epsilon_\mathrm{out}$ and $\epsilon_\mathrm{in}$ of Eq.~\eqref{eq:BinInfidelity}.
		\label{fig:BinIdentification}
        }
\end{figure}

Choose two disjoint frequency intervals, $\Omega_{sL}$ and $\Omega_{sH}$, in each waveguide $s \in \{A,B\}$, representing the lower- and higher-frequency bins, respectively.
Let $|L\rangle_s$ and $|H\rangle_s$ denote single-photon states with normalized spectral amplitudes $\ell_s(\omega)$ and $h_s(\omega)$ supported within the corresponding intervals with $|L\rangle_s = \int_{\Omega_{sL}} \dd\omega\, \ell_s(\omega)\, \hat w_{s\omega}^{\dagger}|0\rangle_s$ and an analogous definition for $|H\rangle_s$, where $\int_{\Omega_{sL}} \dd\omega\, |\ell_s(\omega)|^2 = \int_{\Omega_{sH}} \dd\omega\, |h_s(\omega)|^2 = 1$.
Disjoint supports ensure that the states $|L\rangle_s$ and $|H\rangle_s$ do not overlap, $_s\langle L|H\rangle_s=0$.
The ideal target state now is a \emph{frequency-bin} Bell state defined as
\begin{equation} \label{eq:FrequencyBinBellState}
 |\BellState_{\mathrm{bin}}\rangle = \frac{|L\rangle_A|H\rangle_B + \ee^{\ii \varphi}|H\rangle_A|L\rangle_B}{\sqrt{2}}, 
\end{equation}
where $\varphi$ denotes the relative phase between the two pair-emission alternatives.
Analogously to Eq.~\eqref{eq:BellFidelity}, we define the fidelity of the emitted state with the best-matching frequency-bin Bell state
\begin{equation} \label{eq:FrequencyBinFidelityDef}
	F_{\mathrm{bin}} = \max_{\tilde{\bm{\beta}}} \left|\langle \BellState_{\mathrm{bin}}|\Psi_{\mathrm{out}}\rangle\right|^2,
\end{equation}
with the maximization running over the four mode envelopes and the phase at \emph{fixed} bins, $\tilde{\bm{\beta}} = \{\ell_A,h_A,\ell_B,h_B,\varphi\}$.

To evaluate the fidelity given by Eq.~\eqref{eq:FrequencyBinFidelityDef}, we restrict the output JSA $\Phi_{\mathrm{out}}(\omega_A,\omega_B)$ to a pair of cross-bin rectangles 
\begin{equation}
\begin{split} \label{eq:BinBlocks}
    \Phi_{LH}(\omega_A,\omega_B) &= \bm{i}_{\Omega_{AL}}(\omega_A)\, \bm{i}_{\Omega_{BH}}(\omega_B)\, \Phi_{\mathrm{out}}(\omega_A,\omega_B),
    \\
    \Phi_{HL}(\omega_A,\omega_B) &= \bm{i}_{\Omega_{AH}}(\omega_A)\, \bm{i}_{\Omega_{BL}}(\omega_B)\, \Phi_{\mathrm{out}}(\omega_A,\omega_B).
\end{split}
\end{equation}
Here, $\bm{i}_{\Omega}(\omega)$ denotes the indicator function of the interval $\Omega$, which equals unity for $\omega\in\Omega$ and zero otherwise.
Multiplication by the two indicator functions thus restricts the JSA to the corresponding frequency-bin rectangle.
Note that, unlike the full output JSA $\Phi_{\mathrm{out}}(\omega_A,\omega_B)$, the restricted JSAs $\Phi_{LH}(\omega_A,\omega_B)$ and $\Phi_{HL}(\omega_A,\omega_B)$ are \emph{not normalized}.
The probability of finding the photons in the cross bins is then determined as $P_\mathrm{cross} = P_{LH} + P_{HL}$, where $P_{bb'} = \int \dd\omega_A \int \dd\omega_B \, |\Phi_{bb'}(\omega_A,\omega_B)|^2$ is the probability of finding one photon in bin $b$ and the other photon in bin $b'$ with $b,b' \in \{L,H\}$.
Each restricted JSA possesses its own Schmidt decomposition.
The largest overlap of each restricted JSA with a normalized product mode equals its leading Schmidt amplitude, $\sqrt{\mu_{LH}}$ or $\sqrt{\mu_{HL}}$.
Since the two rectangles involve distinct local modes with disjoint supports, these maxima can be attained simultaneously.
Optimization of the relative phase $\varphi$ then makes the two overlap contributions add constructively.
Hence, the maximal fidelity with a bin-constrained Bell state $|\BellState_\mathrm{bin}\rangle$ is
\begin{equation} \label{eq:FrequencyBinBellFidelity}
 \Fent_\mathrm{bin} = \frac{1}{2}\bigl(\sqrt{\mu_{LH}}+\sqrt{\mu_{HL}}\bigr)^2.
\end{equation}
Since the bin-constrained target family is contained in the unrestricted one and is supported on the two rectangles, $\Fent_\mathrm{bin} \leq \min(\Fent, P_\mathrm{cross})$.

The infidelity separates into three contributions,
\begin{equation} \label{eq:BinInfidelity}
  1 - \Fent_\mathrm{bin}
		= \underbrace{1-P_\mathrm{cross}}_{\epsilon_\mathrm{out}}
        + \underbrace{P_\mathrm{cross}-\mu_{LH}-\mu_{HL}}_{\epsilon_\mathrm{in}}
		+ \underbrace{\tfrac{1}{2}\bigl(\sqrt{\mu_{LH}}-\sqrt{\mu_{HL}}\bigr)^2}_{\epsilon_\mathrm{imb}} ,
\end{equation}
which diagnose, respectively, emission into the same-bin rectangles or outside the bins altogether, residual multimode structure \emph{inside} each cross-bin rectangle, and imbalance between the leading Schmidt contributions of the two cross-bin rectangles.

We evaluate Eqs.~\eqref{eq:FrequencyBinBellFidelity} and~\eqref{eq:BinInfidelity} over the same plane $(g/\kappa,\chi/\kappa)$ of dimensionless coupling rates as in Fig.~\ref{fig:BellFidelity}.
As illustrated in Fig.~\ref{fig:BinIdentification}(a), we set the frequency bins to be the two spectral half-axes separated at $\omega_A=\omega_{\res A}$ and $\omega_B = \omega_{\res B}$. 
At the operating point marked by $\star$ in Fig.~\ref{fig:BellFidelity}(b), this yields $P_\mathrm{cross}=0.991$ and $\Fent_\mathrm{bin} = 0.949$, against $\Fent = 0.956$.
Hence, we conclude that the state of the emitted photon pair is close to a frequency-bin Bell state.
The computations reveal that the imbalance term vanishes within numerical precision throughout the $(g/\kappa,\chi/\kappa)$ plane.
This is to be expected as the symmetric choice of the setup parameters ($g_A=g_B$, $\chi_{AB}=\chi_{BA}$, $\kappa_A = \kappa_B$, and $\Delta_A=\Delta_B$) makes the two rectangles exchange-symmetric and hence equally weighted.

The two fidelities do not coincide, and the way they converge identifies the role of the cross-Kerr coupling.
Along the ridge $g_*(\chi)$ of Fig.~\ref{fig:BellFidelity}(a), which for $\chi/\kappa\lesssim35$ continues below the displayed window, reaching $g_*/\kappa\approx2.0$ at $\chi/\kappa=8$, the unrestricted fidelity already reaches $\Fent=0.937$ at $\chi/\kappa=8$, whereas $\Fent_\mathrm{bin}=0.846$ there.
This discrepancy indicates that the output state is close to a Bell pair in \emph{some} pair of spectral modes, however, those modes are not yet confined to separated frequency bins, and more than $1/10$ of the emission spectrum ($\epsilon_\mathrm{out}=0.123$) falls into the same-bin rectangles.
As $\chi$ grows, the two spectral lobes separate, $\epsilon_\mathrm{out}$ falls by more than an order of magnitude to $8.9\times10^{-3}$ at $\chi/\kappa=80$, and the gap $\Fent-\Fent_\mathrm{bin}$ closes from $0.091$ to $0.007$ as shown in Fig.~\ref{fig:BinIdentification}(b).
The remaining term $\epsilon_\mathrm{in}$ varies only weakly, from $0.031$ to $0.042$, and sets the floor: it measures the residual multimode structure of each individual lobe.
The two terms cross near $\chi/\kappa\approx 22$, so that $\epsilon_\mathrm{out}$ dominates the bin-constrained infidelity at weaker cross-Kerr coupling, while $\epsilon_\mathrm{in}$ becomes dominant for larger cross-Kerr coupling rates.
Along the optimized branch, increasing the cross-Kerr coupling is accompanied by greater separation of the spectral lobes and reduced population in the same-bin rectangles.
Consequently, the bin-constrained fidelity approaches the unrestricted Bell-state fidelity.
At larger coupling strengths, the remaining frequency-bin infidelity is dominated by residual multimode structure within the cross-bin rectangles.
This analysis demonstrates how adjustment of the source parameters yields entanglement that can be encoded in separated frequency bins, a feature that the unrestricted fidelity alone does not establish.

\section{Summary and Outlook}
\label{sec:Summary}

In summary, we have proposed and analyzed an on-demand source of frequency-entangled microwave photon pairs based on the qubit–resonator dimer---two transmon--resonator subsystems coupled by cross-Kerr interaction.
Starting from one excitation in each transmon, the source emits one photon into each of two separate waveguides.
The cross-Kerr interaction correlates the emission dynamics through population-dependent frequency shifts while preserving the excitation number in each subsystem.
Importantly, the photons are routed into distinct output channels without requiring frequency matching between the subsystems.
We outline a possible superconducting circuit-QED implementation of the proposed scheme based on Josephson nonlinear couplers.

Our analysis demonstrates that the setup parameters control both the spectral correlations and the modal structure of the emitted entanglement.
The resulting states exhibit distinct joint spectral intensities and Schmidt spectra, including regimes where entanglement is distributed over many modes as well as regimes dominated by two nearly equally weighted Schmidt contributions.
By optimizing the transmon–resonator coupling and a detuning, we obtain fidelity $\Fent \approx 0.956$ with a maximally entangled two-dimensional state (Bell state).
Restricting the target modes to separated frequency bins reduces the fidelity only to $\Fent_\mathrm{bin} \approx 0.949$, so that the emitted pair closely approximates a frequency-bin Bell state.
We identified that in this regime, the residual infidelity is primarily associated with population outside the dominant two-mode sector rather than from the imbalance between its two contributions.
Combined with the well-separated spectral features of the emitted photons, these results motivate the use of the proposed source for frequency-bin entanglement generation and its distribution through distinct microwave channels.

The proposed source of frequency-bin-entangled photon pairs can be utilized for the \emph{conditional} preparation of an itinerant frequency-bin qubit.
The potential preparation protocol is as follows: one of the emitted photons serves as the output frequency-bin qubit, while the other acts as an ancilla, which undergoes a coherent frequency-bin measurement.
The outcome of this measurement heralds the preparation of the output photon in a target superposition of frequency modes.
Essentially, this protocol is a discrete-frequency-bin version of entanglement-assisted single-photon shaping considered in Ref.~\cite{averchenko2017pra}.

Coherent capture of the output photons by separate receiver resonator modes could transfer the photon entanglement to stationary memories, building on microwave state-transfer and remote-entanglement protocols demonstrated in experiment~\cite{kurpiers2018nat}.
Dual-rail resonator encodings are a natural candidate~\cite{teoh2023pnas}, with experimental implementations of erasure detection that distinguish photon loss without resolving the stored logical state~\cite{koottandavida2024prl, degraaf2025npj}.
Two copies of the proposed biphoton source could enable entanglement swapping.
A joint Bell measurement on one of the photons from each pair would herald entanglement between the remaining photons, which can then be transferred to stationary memories.
Coherent capture into superconducting qubits could enable complete Bell measurements using local gates and qubit readout, as demonstrated in superconducting teleportation experiments~\cite{steffen2013nat}.
These capabilities would provide resources for teleportation, entanglement-based communication, and modular quantum processing.

The frequency-bin-entangled photons propagating in distinct waveguides could also be used for phase and delay sensing in circuit QED architecture.
For a frequency anticorrelated pair with equal frequency-bin separations in the two waveguides, differential propagation delays are encoded in the relative phase between the pair-emission alternatives, whereas a common delay leaves the encoded relative phase unchanged.
Local frequency-bin rotations followed by correlated detection would allow this phase to be measured, providing a route to characterizing differential delays or spectral phase shifts in microwave channels.

\begin{acknowledgments}
 The authors thank A.~Semenov for useful comments.
 E.V.S. thanks A.~Sokolov and R.~Baskov for fruitful discussions on various aspects of circuit QED.
 This work was supported by the National Research Foundation of Ukraine through Project No.~2023.03/0165, Quantum correlations of electromagnetic radiation.
 E.V.S. acknowledges partial support from the Simons Foundation in the framework of the call ``Centers of Excellence in Ukraine'', Grant No. SFI-PD-COE-00032628.
\end{acknowledgments}

\subsection*{Data availability statement}
\noindent The data that support the findings of this study are available from the authors upon reasonable request.
%

\appendix


\section{Derivation of the transmon--resonator dimer Hamiltonian}
\label{app:CrossKerr}
\begin{figure}[t!] 
	\centering
	\includegraphics{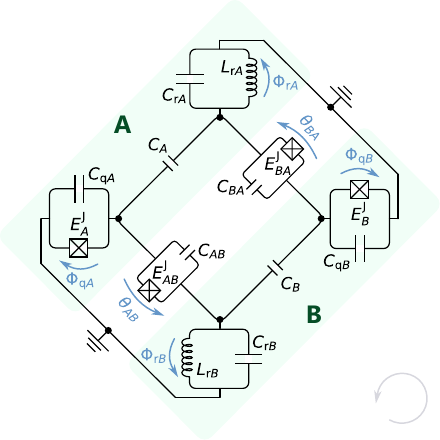}
	\caption{Circuit diagram of the transmon--resonator dimer---two transmon--resonator systems coupled via cross-Kerr couplings.
    Arrows indicate the direction of the flux drop. The loops are traversed counterclockwise. \label{fig:CircuitScheme}}
\end{figure}

Here, we derive the quantum Hamiltonian~\eqref{eq:TransmonsHamiltonian} of the transmon--resonator dimer described in detail in Sec.~\ref{sec:Principle}.
The circuit diagram of this system is shown in Fig.~\ref{fig:CircuitScheme}.

Following the standard procedure of circuit quantization\footnote{Refer, for example, to the reviews by Vool and Devoret~\cite{vool2017} and Rasmussen \textit{et al.}~\cite{rasmussen2021} and references therein.}, we start by writing down the circuit Lagrangian $\Lagr_\mathrm{qr}$, which is expressed as
\begin{equation} \label{eq:Lagrangian}
  \begin{split}
    \Lagr_{\qb\res} = & \, \sum_{s} \left[\frac{C_{\res s} \dot\Phi^2_{\res s}}{2} - \frac{\Phi^2_{\res s}}{2L_{\res s}}\right]
    + \sum_{s}\left[\frac{C_{\qb s} \dot\Phi^2_{\qb s}}{2} + E_{\JJ s}\cos\left(2\pi \frac{\Phi_{\qb s}}{\Phi_0}\right)\right]
    \\
    & \, + \sum_{s} \frac{C_s (\dot\Phi_{\res s} - \dot\Phi_{\qb s})^2}{2} + \sum_{s} \frac{C_{s\bar{s}} \dot\theta_{s\bar{s}}^2}{2}
    + \sum_{s} E_{\JJ s\bar{s}} \cos\left(2\pi \frac{\theta_{s\bar{s}}}{\Phi_0}\right),
  \end{split}
\end{equation}
where $\Phi_0 = h/2e$ denotes the flux quantum. 
In the upper line, the first term describes the resonators, whereas the second term corresponds to the transmons (qubits).
The bottom line describes couplings within and between the subsystems $A$ and $B$.
Therein, the first term corresponds to the capacitive coupling of the resonator to the transmon within the same subsystem.
The second term describes the capacitive coupling between the different subsystems, mediated by the capacitances of the couplers.
The total capacitance of the coupler is formed by the parallel connection of the self-capacitance of its JJ and the (optional) shunting capacitance.
Finally, the third term describes the inductive coupling between the subsystems due to the Josephson nonlinear inductance.
In Eq.~\eqref{eq:Lagrangian}, $\Phi_{\res s}$ and $\Phi_{\qb s}$ denote the flux drop over the resonator and the qubit in the subsystem $s$, respectively, whereas $\theta_{s\bar{s}}$ stands for the flux drop over the coupler connecting the qubit in the subsystem $s$ and the resonator in the other subsystem $\bar{s}$.
The rest of the notations introduced in the Lagrangian~\eqref{eq:Lagrangian} are listed in Table~\ref{tab:CircuitNotations}.

\begin{table}[t!]
	\centering
	\caption{Notations for the Lagrangian parameters.
    }
	\label{tab:CircuitNotations}
	
	\begin{tabular}{l p{0.45\textwidth}}
		\toprule
		Parameter
		& Description \\
		\midrule
		$L_{\res s}$	& Resonator inductance
        \\
		$C_{\res s}$	& Resonator capacitance
        \\
		$E_s^{\JJ}$		& Josephson energy of the transmon junction
        \\
		$C_{\qb s}$		& Transmon capacitance
        \\
		$C_{s}$ 		& Capacitance coupling the transmon to the resonator mode
        \\
		$E_{s\bar{s}}^{\JJ}$	& Josephson energy of the junction in the coupler bridging the transmon $s$ and the resonator $\bar{s}$.
        \\
		$C_{s\bar{s}}$ 		& Total capacitance of the coupler that connects the transmon $s$ and the resonator $\bar{s}$ 
        \\
		\bottomrule
	\end{tabular}
\end{table}

Using the flux quantization rule (see, e.g., Ref.~\cite{barone1982}), we can express the flux drops over the couplers as $\theta_{AB} = \Phi_{\qb A} - \Phi_{\res B}$ and $\theta_{BA} = \Phi_{\qb B} - \Phi_{\res A}$.
Substituting these relationships into Eq.~\eqref{eq:Lagrangian} and rearranging the terms leads to the following result
\begin{equation} \label{eq:Lagrangian2}
  \begin{split}
    \Lagr_{\qb\res} = & \, \sum_{s} \left[\frac{C'_{\res s} \dot\Phi^2_{\res s}}{2} - \frac{\Phi^2_{\res s}}{2L_{\res s}}\right]
    + \sum_{s}\left[\frac{C'_{\qb s} \dot\Phi^2_{\qb s}}{2} + E_s^{\JJ}\cos\left(2\pi \frac{\Phi_{\qb s}}{\Phi_0}\right)\right]
    \\
    & \, - \sum_s (C_s \dot\Phi_{\res s} \dot\Phi_{\qb s} + C_{s\bar{s}} \dot\Phi_{\qb s} \dot\Phi_{\res \bar{s}})
    + \sum_{s} E_{s\bar{s}}^{\JJ} \cos\left(2\pi \frac{\Phi_{\qb s} - \Phi_{\res \bar{s}}}{\Phi_0}\right),
  \end{split}
\end{equation}
where $C'_{\res s} = C_{\res s} + C_s + C_{\bar{s}s}$ and $C'_{\qb s} = C_{\qb s} + C_s + C_{s\bar{s}}$.

Next, by applying the Legendre transformation~\cite{goldstein1980}, one derives the classical Hamiltonian of the transmon--resonator dimer:
\begin{equation} \label{eq:ClassicalHamitonian}
\begin{split}
        H_{\qb\res} = & \sum_{s} \sum_{\mu=\mathrm{q},\mathrm{r}} Q_{\mu s} \dot \Phi_{\mu s} \, - \Lagr_\mathrm{qr},
\end{split}
\end{equation}
where $Q_{\mu s} = \partial \Lagr_{\qb\res}/\partial \dot\Phi_{\mu s}$ is the generalized momentum, which corresponds to the charge accumulated on the capacitance of the resonator ($\mu = \res$) or the qubit ($\mu =\qb$) in the subsystem $s$.
The generalized momenta (charges) are determined as $\bm{Q} = \mathsf{C} \dot{\bm{\Phi}}$, where $\bm{Q} = (Q_{\res A}, Q_{\res B}, Q_{\qb A}, Q_{\qb B})^T$ and $\bm{\Phi} = (\Phi_{\res A},  \Phi_{\res B}, \Phi_{\qb A}, \Phi_{\qb B})^T$.
We formally represent the capacitance matrix $\mathsf C$ as
\begin{equation} \label{eq:CapacitanceMatrix}
	\mathsf C = \mathsf D-\mathsf V,
\end{equation}
where
\begin{equation}
	\mathsf D
	=
    \begin{pmatrix}
		C'_{\res A} & 0 & 0 & 0\\
		0 & C'_{\res B} & 0 & 0\\
		0 & 0 & C'_{\qb A} & 0\\
		0 & 0 & 0 & C'_{\qb B}
	\end{pmatrix},
	\qquad
	\mathsf V
	=
	\begin{pmatrix}
		0 & 0 & C_A & C_{BA}\\
		0 & 0 & C_{AB} & C_B\\
		C_A & C_{AB} & 0 & 0\\
		C_{BA} & C_B & 0 & 0
	\end{pmatrix}.
\end{equation}

Assuming that the coupling capacitances are small compared with the node capacitances, the inverse capacitance matrix can then be approximately expanded as $\mathsf C^{-1} \simeq \mathsf D^{-1} + \mathsf D^{-1}\mathsf V\mathsf D^{-1}$.
Thus, to the retained order, $(\mathsf C^{-1})_{\mu s,\mu s} \simeq 1/{C'_{\mu s}}$, $(\mathsf C^{-1})_{\qb s,\res s} \simeq {C_s}/{C'_{\qb s}C'_{\res s}}$, and $	(\mathsf C^{-1})_{\qb s, \res \bar s} \simeq {C_{s\bar{s}}}/{C'_{\qb s}C'_{\res \bar{s}}}$.
The qubit--qubit and resonator--resonator entries of $\mathsf C^{-1}$ first appear at second order in this expansion and are neglected.

Writing the Lagrangian as $\mathcal L_{qr} =\dot{\bm\Phi}^{\,T} \mathsf C \dot{\bm\Phi}/2 - U(\bm\Phi)$, the Legendre transformation leads to the Hamiltonian $H_{\qb\res} = T(\bm Q) + U(\bm{\Phi})$, where
\begin{equation}
    T(\bm Q) = \frac{1}{2}\bm{Q}^T\mathsf C^{-1}\bm{Q}, \quad U(\boldsymbol\Phi)
	= \sum_s\frac{\Phi_{rs}^2}{2L_{\res s}} - \sum_s E_s^\JJ \cos\left(2\pi\frac{\Phi_{\qb s}}{\Phi_0}\right) - \sum_s E_{s\bar s}^\JJ
	\cos\left[2\pi\frac{(\Phi_{\qb s}-\Phi_{\res \bar s})}{\Phi_0}\right].
\end{equation}
Using the first-order approximation to $\mathsf C^{-1}$, the ``kinetic energy'' part $T(\bm Q)$ becomes
\begin{equation}
	T(\bm Q) \simeq
	\sum_s \left(\frac{Q_{\res s}^2}{2C'_{\res s}} + \frac{Q_{\qb s}^2}{2C'_{\qb s}}\right)
	+ \sum_s \frac{C_s}{C'_{\qb s}C'_{\res s}} Q_{\qb s}Q_{\res s} + \sum_s \frac{C_{s\bar s}} {C'_{\qb s}C'_{\res \bar s}} Q_{\qb s}Q_{\res \bar s}.
\end{equation}
Next, we introduce the reduced charge and flux variables $n_{\mu s} = Q_{\mu s}/2e$ and $\phi_{\mu s} = 2\pi \Phi_{\mu s}/\Phi_0$.
The former can be interpreted as a number of excess (or deficient) Cooper pairs on the element $\mu$ in the subsystem $s$, while the latter is the gauge-invariant condensate phase drop across this element.
In terms of these variables, the Hamiltonian reads
\begin{equation} \label{eq:ClassicalHamiltonian2}
 \begin{split}
    H_{\qb\res} = & \, \sum_{s} \Big(4E^C_{\res s} n_{\res s}^2 + E_{\res s}^{L}\frac{\phi_{\res s}^2}{2}\Big)
    + \sum_{s} \Big(4E_{\qb s}^{C} n_{\qb s}^2 - E_s^{\JJ} \cos \phi_{\qb s}\Big)
    \\
    & \, + \sum_{s} E^{C}_s n_{\qb s} n_{\res s} + \sum_{s} E^C_{s\bar{s}} n_{\qb s} n_{\res \bar{s}} - \sum_{s} E_{s\bar{s}}^{\JJ} \cos (\phi_{\qb s} - \phi_{\res \bar{s}}).
 \end{split} 
\end{equation}
In the Hamiltonian~\eqref{eq:ClassicalHamiltonian2}, $E^L_{\res s} = L_{\res s}^{-1} (\Phi_0/2\pi)^2$ is the inductive energy of the resonator $s$.
Within the adopted approximation, the capacitive energies of the resonator and the qubit are $E^C_{\res s} = e^2/2 C'_{\res s}$ and $E^C_{\qb s} = e^2/2 C'_{\qb s}$, respectively, while the capacitive interaction energies are $	E_s^C = {4e^2 C_s}/{C'_{\qb s}C'_{\res s}}$ and $E_{s\bar s}^C = {4e^2 C_{s\bar s}}/{C'_{\qb s}C'_{\res \bar s}}$.

By promoting the reduced charge and phase (flux) variables to the quantum operators that obey the commutation relations $[\hat \phi_{\mu s}, \hat n_{\mu' s'}] = \ii \delta_{\mu \mu'} \delta_{s s'}$, we obtain the quantized Hamiltonian of the transmon--resonator dimer.
Expanding the cosine terms into a Taylor series, we arrive at the Hamiltonian as follows
\begin{equation} \label{eq:QuantizedHamiltonian}
 \begin{split}
    \hat H_{\qb\res} = & \, \sum_{s} \Big(4E_{\res s}^C \hat n_{\res s}^2 + \tilde{E}_{\res s}^L\frac{\hat\phi_{\res s}^2}{2}\Big)
    + \sum_{s} \Big[4E_{\qb s}^C \hat n_{\qb s}^2 + \tilde E_s^{\JJ}\frac{\hat\phi_{\qb s}^2}{2} -  E^\JJ_{s} \sum_{k\geq 2} \frac{(-1)^k}{(2k)!}\hat\phi_{\qb s}^{2k}\Big] \\
    & \, + \sum_s E^C_s \hat n_{\qb s} \hat n_{\res s} + \sum_{s} E_{s\bar{s}}^C \hat n_{\qb s} \hat n_{\res \bar{s}}
    - \sum_s E^\JJ_{s\bar{s}} \hat\phi_{\qb s} \hat\phi_{\res\bar{s}}
    - \sum_{s} E_{s\bar{s}}^{\JJ} \sum_{k\geq 2} \frac{(-1)^k}{(2k)!} (\hat\phi_{\qb s} - \hat\phi_{\res \bar{s}})^{2k},
 \end{split}  
\end{equation}
where $\tilde E^L_{\res s} = E^L_{\res s} + E_{\bar{s}s}^{\JJ}$ is the renormalized inductive energy of the resonator $s$, whereas $\tilde E^\JJ_s = E^\JJ_s + E^\JJ_{s\bar{s}}$ stands for the renormalized Josephson energy of the transmon $s$.

Next, we express the charge and flux operators of the resonator $s$ in terms of the bosonic ladder operators as $\hat n_{\res s} = \ii n^\mathrm{zpf}_{\res s} (\hat a^\dag_{s} - \hat a_s)$ and $\hat \phi_{\res s} = \varphi^\mathrm{zpf}_{\res s}(\hat a^\dag_s + \hat a_s)$, where operator $\hat a_s$ annihilates a photon in a mode of the resonator in subsystem $s$.
Analogously, the operators of the qubit $s$ are represented as $\hat n_{\qb s} = \ii n^\mathrm{zpf}_{\qb s}(\hat b^\dag_s - \hat b_s)$ and $\hat \phi_{\qb s} = \varphi^\mathrm{zpf}_{\qb s} (\hat b^\dag_s + \hat b_s)$, with $\hat b_s$ being the annihilation operator of an excitation in the qubit in subsystem $s$.
Here, $n^\mathrm{zpf}_{\res s} = (\tilde E^L_{\res s}/32E^C_{\res s})^{1/4}$ and $\varphi^\mathrm{zpf}_{\res s} = (2E^C_{\res s}/\tilde E^L_{\res s})^{1/4}$ are the magnitudes of the charge and phase zero-point fluctuations of the resonator $s$.
The magnitudes of the zero point fluctuations of the qubit $s$ are denoted as $n^\mathrm{zpf}_{\qb s} = (\tilde E^\JJ_{s}/32E^C_{\qb s})^{1/4}$ and $\varphi^\mathrm{zpf}_{\qb s} = (2E^C_{\qb s}/\tilde E^\JJ_s)^{1/4}$.

We substitute the above ladder-operator representations of the resonator and transmon charge and phase operators into Eq.~\eqref{eq:QuantizedHamiltonian}.
In the low-excitation regime considered here, assuming small zero-point phase fluctuations,
$(\varphi^{\mathrm{zpf}}_{\res s})^2, (\varphi^{\mathrm{zpf}}_{\qb s})^2 \ll 1$, and retaining only the leading (quartic) nonlinear terms, we obtain\footnote{A generalization to a qubit with both cubic and quartic nonlinearities coupled to a resonator via a flux-biased symmetric SQUID is discussed in detail in Ref.~\cite{stolyarov2026pra}.}
\begin{equation} \label{eq:QuantizedHamiltonian2}
 \begin{split}
    \frac{\hat H_{\qb\res}}{\hbar} \approx & \, \sum_s \left[\omega^p_{\res s} \hat a^\dag_s \hat a_s - Y_{\res s}(\hat a^\dag_s + \hat a_s)^4\right]
    + \sum_{s} \left[\omega^p_{\qb s} \hat b^\dag_s \hat b_s - Y_{\qb s}(\hat b^\dag_s + \hat b_s)^4\right]
    \\
    & \, - \sum_s g_s (\hat a^\dag_s - \hat a_s)(\hat b^\dag_s - \hat b_s)
    - \sum_{s} \left[g^c_{s\bar{s}} (\hat a^\dag_{\bar{s}} - \hat a_{\bar{s}})(\hat b^\dag_s - \hat b_s) + g^i_{s\bar{s}}(\hat a^\dag_{\bar{s}} + \hat a_{\bar{s}})(\hat b^\dag_{s} + \hat b_{s})\right] 
    \\
    & \, - \sum_s \frac{\chi_{s\bar{s}}}{4}(\hat a^\dag_{\bar{s}} + \hat a_{\bar{s}})^2 (\hat b^\dag_{s} + \hat b_{s})^2 + \sum_s g^{(13)}_{s\bar{s}} (\hat b^\dag_{s} + \hat b_{s})(\hat a^\dag_{\bar{s}} + \hat a_{\bar{s}})^3 + \sum_s g^{(31)}_{s\bar{s}} (\hat b^\dag_{s} + \hat b_{s})^3(\hat a^\dag_{\bar{s}} + \hat a_{\bar{s}}).
 \end{split}
\end{equation}
The terms in the upper line describe the resonator and the transmon as nonlinear oscillators with (plasma) frequencies $\omega^p_{\res s} = \sqrt{8 E^C_{\res s} \tilde{E}^L_{\res s}}/\hbar$ and $\omega^p_{\qb s} = \sqrt{8 E^C_{\qb s} \tilde E^\JJ_{s}}/\hbar$, and quartic self-nonlinearities of strengths $Y_{\res s} = \frac{1}{24\hbar} E^\JJ_{\bar{s}s} (\varphi^\mathrm{zpf}_{\res s})^4$ and $Y_{\qb s} = \frac{1}{24\hbar} \tilde E^\JJ_s (\varphi^\mathrm{zpf}_{\qb s})^4$, respectively.

The second line in the Hamiltonian~\eqref{eq:QuantizedHamiltonian2} describes various \emph{linear} couplings in the transmon--resonator dimer.
Therein, the first term corresponds to the linear interaction between the transmon and the resonator within \emph{the same subsystem}.
This interaction is of purely capacitive origin, and its rate is given by
\begin{equation} \label{eq:LinearCouplingCapacitive}
    g_s = \frac{E^C_{s}}{\hbar} n^\mathrm{zpf}_{\res s} n^\mathrm{zpf}_{\qb s}.
\end{equation}
The next term describes linear couplings between the transmon and the resonator belonging to \emph{different subsystems}, mediated by the JJ coupler.
The first term in the brackets corresponds to the capacitive-type interaction between the transmon $s$ and the resonator $\bar{s}$.
This interaction is mediated by the parallel-connected shunting capacitance of the coupler and the self-capacitance of its JJ.
The second term in the brackets is the inductive coupling mediated solely by the JJ.
The rates of these couplings are~\cite{stolyarov2025detector, stolyarov2026pra}
\begin{equation} \label{eq:LinearCrossCouplings}
    g^c_{s\bar{s}} = \frac{E^C_{s\bar{s}}}{\hbar} n^\mathrm{zpf}_{\qb s} n^\mathrm{zpf}_{\res \bar{s}}, \quad g^i_{s\bar{s}} = \frac{E^\JJ_{s\bar{s}}}{\hbar} \varphi^\mathrm{zpf}_{\qb s} \varphi^\mathrm{zpf}_{\res \bar{s}}.
\end{equation}

The bottom line in the Hamiltonian~\eqref{eq:QuantizedHamiltonian2} describes \emph{nonlinear} interactions between the subsystems $A$ and $B$.
These couplings arise due to nonlinearity of the JJ couplers.
Therein, the first term gives rise to the cross-Kerr coupling between the transmon $s$ and the resonator $\bar{s}$ with the rate
\begin{equation} \label{eq:CrossKerrCoupling}
    \chi_{s\bar{s}} = \frac{E^\JJ_{s\bar{s}}}{\hbar} (\varphi^\mathrm{zpf}_{\qb s} \varphi^\mathrm{zpf}_{\res \bar{s}})^2.
\end{equation}
The remaining terms lead, among others, to processes of conversion of one excitation in one subsystem into three excitations in the other subsystem with the rates
\begin{equation} \label{eq:ThreeExcitation}
    g^{(31)}_{s\bar{s}} = \frac{E^\JJ_{s\bar{s}}}{6\hbar} (\varphi^\mathrm{zpf}_{\qb s})^3 \varphi^\mathrm{zpf}_{\res\bar{s}}, \quad g^{(13)}_{s\bar{s}} = \frac{E^\JJ_{s\bar{s}}}{6\hbar} \varphi^\mathrm{zpf}_{\qb s} (\varphi^\mathrm{zpf}_{\res\bar{s}})^3.
\end{equation}

To proceed, we outline the approximations that reduce Eq.~\eqref{eq:QuantizedHamiltonian2} to the effective Hamiltonian presented below.
We first consider the interaction picture with respect to the harmonic part of the Hamiltonian~\eqref{eq:QuantizedHamiltonian2}.
We neglect the counter-rotating terms, which oscillate at sums of oscillator frequencies.
Next, we assume that the setup operates away from the corresponding multiphoton resonances, with the relevant transition detunings much larger than the coupling matrix elements.
Within the RWA, we neglect the corresponding terms.
However, this approximation does not eliminate all nonlinear exchange processes.
In particular, the terms proportional to $g_{s\bar s}^{(31)}$ and $g_{s\bar s}^{(13)}$ produce both a correction to the single-excitation exchange coupling and the occupation-dependent single-photon exchange terms.
Moreover, the first term in the bottom line of Eq.~\eqref{eq:QuantizedHamiltonian2} generates pair exchange process $\propto (\hat a^{\dag 2}_{\bar s} \hat b^{2}_{s} + \hat b^{\dag 2}_{s} \hat a_{\bar{s}}^2)$ in addition to the diagonal cross-Kerr interaction.

We retain the correction to the single-excitation exchange coupling.
However, we assume that the remaining nonlinear exchange processes are sufficiently far off resonance to be neglected.
Specifically, their matrix elements must be small compared with the corresponding transition detunings, including the anharmonic and cross-Kerr shifts.
Higher-order dispersive corrections generated by these processes are assumed negligible at the accuracy considered here.
This regime can be achieved by appropriately separating the operating frequencies of subsystems $A$ and $B$, while maintaining the near-resonant regime of interaction between the transmon and the resonator within each subsystem.
Under these conditions, the Hamiltonian~\eqref{eq:QuantizedHamiltonian2} reduces to
\begin{equation} \label{eq:QuantizedHamiltonianRWA}
 \begin{split}
    \frac{\hat H'_{\qb\res}}{\hbar} \approx & \, \sum_s \left(\omega_{\res s} \hat a^\dag_s \hat a_s - \frac{K_{\res s}}{2} \hat a^{\dag 2}_s \hat a^2_s\right) + \sum_{s} \left(\omega_{\qb s} \hat b^\dag_s \hat b_s - \frac{K_{\qb s}}{2}\hat b^{\dag 2}_s \hat b^2_s\right)
    \\
    & \, + \sum_s g_s (\hat a^\dag_s \hat b_s + \hat b^\dag_s \hat a_s)
    + \sum_s g_{s\bar{s}} (\hat a^\dag_{\bar{s}} \hat b_{s} + \hat b^\dag_{s} \hat a_{\bar{s}}) - \sum_s \chi_{s\bar{s}} \hat a^\dag_{\bar{s}} \hat a_{\bar{s}} \hat b^\dag_s \hat b_s.
 \end{split}
\end{equation}
Here, the oscillator frequencies include the frequency shifts generated by the Josephson nonlinearities: $\omega_{\res s} = \omega^p_{\res s} - K_{\res s} - \chi_{\bar{s} s}/2$ and $\omega_{\qb s} = \omega^p_{\qb s} - K_{\qb s} - \chi_{s \bar{s}}/2$.
The strengths of the resonator and qubit self-Kerr nonlinearities are $K_{\res s} = 12Y_{\res s}$ and $K_{\qb s} = 12Y_{\qb s}$, respectively.

In the Hamiltonian~\eqref{eq:QuantizedHamiltonianRWA}, the single-excitation exchange rate between the transmon in subsystem $s$ and the resonator in subsystem $\bar{s}$ is  determined as $g_{s\bar{s}} = g^c_{s\bar{s}} - g^i_{s\bar{s}} + 3g^{(13)}_{s\bar{s}} + 3g^{(31)}_{s\bar{s}}$.
Importantly, a proper choice of the shunting capacitance in the coupler allows one to cancel the single-photon coupling~\cite{stolyarov2026pra, neumeier2013, ciani2019qst, sank2025prappl} between the subsystems.
Having suppressed the excitation-exchange processes between different subsystems, we obtain the effective circuit Hamiltonian
\begin{equation} \label{eq:QuantizedHamiltonianEffective}
 \begin{split}
    \frac{\hat H^\mathrm{eff}_{\qb\res}}{\hbar} = & \, \sum_s \left(\omega_{\res s} \hat a^\dag_s \hat a_s - \frac{K_{\res s}}{2} \hat a^{\dag 2}_s \hat a^2_s\right) + \sum_{s} \left(\omega_{\qb s} \hat b^\dag_s \hat b_s - \frac{K_{\qb s}}{2}\hat b^{\dag 2}_s \hat b^2_s\right)
    + \sum_s g_s (\hat a^\dag_s \hat b_s + \hat b^\dag_s \hat a_s) - \sum_s \chi_{s\bar{s}} \hat a^\dag_{\bar{s}} \hat a_{\bar{s}} \hat b^\dag_s \hat b_s.
 \end{split}
\end{equation}
Notably, the Hamiltonian combining the above Hamiltonian and the waveguide Hamiltonian~\eqref{eq:WaveguideHamiltonian} conserves the number of excitations in each subsystem, $\big[\hat {\mathcal{N}}_s, \hat H^\mathrm{eff}_{\qb\res} + \hat H_\wg\big] = 0$, where $\hat{\mathcal{N}}_s = \hat a^\dag_s \hat a_s + \hat b^\dag_s \hat b_s + \int \dd\omega \, \hat w^\dag_{s \omega} \hat w_{s \omega}$ is the operator of the excitation number in the subsystem $s$.
Recall that the initial state of the setup has one excitation in each transmon, with the resonators and waveguides in their
vacuum states.
Consequently, since the number of excitations in each subsystem is conserved, the subsequent evolution remains within the sector that contains a total of two excitations.
Each transmon--resonator subsystem contains at most one excitation, with the remaining excitation probability residing in its corresponding waveguide.
Within this sector, only the two lowest transmon states $|g\rangle_s$ and $|e\rangle_s$ are accessible.
Then, we replace the bosonic operators with the two-level system ladder operators, $\hat b_s \rightarrow \hat \sigma^-_s$.
Furthermore, both self-Kerr operators, $\hat b_s^{\dagger 2}\hat b_s^2$ and $\hat a_s^{\dagger 2}\hat a_s^2$, annihilate every state in
the accessible sector.
Hence, their omission follows from the excitation-number restriction and does not require the self-Kerr nonlinearities to vanish.
Equation~\eqref{eq:QuantizedHamiltonianEffective} consequently reduces, within the accessible sector, to the Hamiltonian~\eqref{eq:TransmonsHamiltonian} in Sec.~\ref{sec:Hamiltonian}.


\section{Details of numerical computations}
\label{app:NumericalMethods}

This appendix summarizes the numerical procedures behind the results of Secs.~\ref{sec:OutputState} and~\ref{sec:FrequencyBinEntanglement}.
The emitted state is obtained by integrating the evolution equations for the probability amplitudes~\eqref{eq:EvolutionEquations}; see details in Appendix~\ref{app:ODEs_Integrals}.
The joint spectral amplitude of the emitted photons is discretized on a frequency grid and decomposed using SVD to obtain the Schmidt spectrum, Schmidt number $K$, and fully-entangled fraction $\Fent$.
This procedure is discussed in Appendix~\ref{app:SVDDetails}.
Finally, $\Fent$ is maximized over $(|g|,\Delta)$ at fixed $\chi$.
The maximization strategy is outlined in Appendix~\ref{app:FidelityOptimization}.
All computations were performed in \textsc{Julia}~\cite{bezanson2017julia}.

\subsection{Time evolution and frequency integrals} 
\label{app:ODEs_Integrals}

We solve the set of ordinary differential equations in Eq.~\eqref{eq:EvolutionEquations} numerically using adaptive Runge-Kutta methods implemented in \textsf{DifferentialEquations.jl} library~\cite{rackauckas2017}.
For the equal-time cross-correlation $g^{(2)}_{AB}(t)$, we instead exploit the linearity of Eq.~\eqref{eq:EvolutionEquations} at fixed waveguide frequencies: the amplitude vector $\mathbf{u}(t;\omega_A,\omega_B) = \exp[\mathbf{M}(\omega_A,\omega_B)\,t] \, \mathbf{u}_0$ is obtained by matrix exponentiation, yielding an exact solution free of temporal discretization error.
The one-dimensional frequency integrals are evaluated using Gauss-Kronrod quadrature~\cite{laurie1997} implemented in \textsf{QuadGK.jl} library~\cite{quadgk}.
The joint two-photon probability Eq.~\eqref{eq:TwoPhotonProbability} requires a two-dimensional integral computed via the $h$-adaptive Genz-Malik cubature rule~\cite{genz1980}.

\subsection{Singular-value decomposition} \label{app:SVDDetails}

The asymptotic JSA is sampled on a uniform square grid,
$\omega_i = -W + (i-1)\delta\omega$ with $i = 1,\dots,n$ and
$n = 2W/\delta\omega + 1$ points per axis, which turns
$\Phi_\mathrm{out}(\omega_A,\omega_B)$ into the matrix
$\mathsf{M}_{ij} = \Phi_\mathrm{out}(\omega_i,\omega_j)$.
The half-width $W$ must enclose the spectral support of the amplitude.
Since the emission acquires sidebands at the Rabi frequency and is displaced by the cross-Kerr interaction, $W$ is scaled as $W \propto |g| + \chi + |\Delta|$, subject to a lower bound of a few $\kappa$ that keeps the window wider than the bare resonator emission linewidth as $g,\chi \to 0$.

On this grid, the kernels~\eqref{eq:IntegralKernels} reduce to the Gram matrices $\mathcal{U} \to \delta\omega\,\mathsf{M}\mathsf{M}^\dag$ and $\mathcal{V} \to \delta\omega\,\mathsf{M}^\dag\mathsf{M}$, so that Eqs.~\eqref{eq:EigenvalueProblem} become the eigenvalue problems of $\mathsf{M}\mathsf{M}^\dag$ and $\mathsf{M}^\dag\mathsf{M}$.
These two matrices share the squared singular values of $\mathsf{M}$ as their spectrum, and it is preferable not to form them explicitly: computing the SVD of $\mathsf{M}$ directly avoids squaring the condition number~\cite{golub2013, trefethen1997}, which matters here because the Schmidt spectrum spans several orders of magnitude once the emitted state becomes strongly multimode.
Writing $\mathsf{M} = \mathsf{U}\,\mathrm{diag}(\sigma_k)\,\mathsf{V}^\dag$ with $\sigma_1 \geq \sigma_2 \geq \dots \geq 0$, where $\mathsf{U}$ and $\mathsf{V}$ are unitary and their columns $\mathsf{u}_k$, $\mathsf{v}_k$ are the left and right singular vectors, which satisfy $\mathsf{M}\mathsf{v}_k = \sigma_k\mathsf{u}_k$ and $\mathsf{M}^\dag\mathsf{u}_k = \sigma_k\mathsf{v}_k$.
The singular values $\sigma_k$ are thus the square roots of the eigenvalues of $\mathsf{M}^\dag\mathsf{M}$ (equivalently, of $\mathsf{M}\mathsf{M}^\dag$), and at most $n$ of them are nonzero.
The Schmidt modes are determined as $\textstyle u_k(\omega_i) = \mathsf{U}_{ik}/\sqrt{\delta\omega}$ and $\textstyle v_k(\omega_i) = \mathsf{V}^*_{ik}/\sqrt{\delta\omega}$.
The Schmidt coefficients are $\textstyle \lambda_k = \delta\omega^2 \sigma^2_k = \sigma^2_k/\sum_l \sigma^2_l$.
The two expressions for $\lambda_k$ agree because the discrete form of the normalization $\textstyle \int\dd\omega_A \int \dd\omega_B |\Phi_\mathrm{out}(\omega_A, \omega_B)|^2 = 1$ is precisely $\textstyle \delta\omega^2 \sum_l \sigma^2_l = 1$.
In practice, we evaluate the populations as the ratio $\textstyle \sigma^2_k / \sum_l \sigma^2_l$ rather than as $\delta\omega^2 \sigma^2_k$:
the ratio enforces $\textstyle \sum_k \lambda_k = 1$ by construction rather than only in the continuum limit, and it is insensitive to any overall scale factor carried by $\Phi_\mathrm{out}$.
The factor $1/\sqrt{\delta\omega}$ appears solely in the modes, which are densities in $\omega$ and satisfy $\textstyle \int\dd\omega\,|u_k(\omega)|^2 = 1$
because the singular vectors are unit vectors.
Their orthonormality likewise reproduces that of the Schmidt modes
automatically, so no subsequent re-orthogonalization is required.

The two discretization parameters control different errors: $W$ sets the
truncation of the spectral tails, whereas $\delta\omega$ determines how well the narrow interference structure of the JSA is resolved.
Rather than fixing $\delta\omega$ in advance, we refine it adaptively---the grid is successively halved until \emph{both} the Schmidt number and the Bell-likeness introduced in Sec.~\ref{sec:FrequencyBinEntanglement} are stable to a prescribed relative tolerance, subject to a floor $\delta\omega_\mathrm{min}$ set by the
available computation time.

\begin{figure*}[t!] 
	\centering
	\hfill\includegraphics[width=0.82\textwidth]{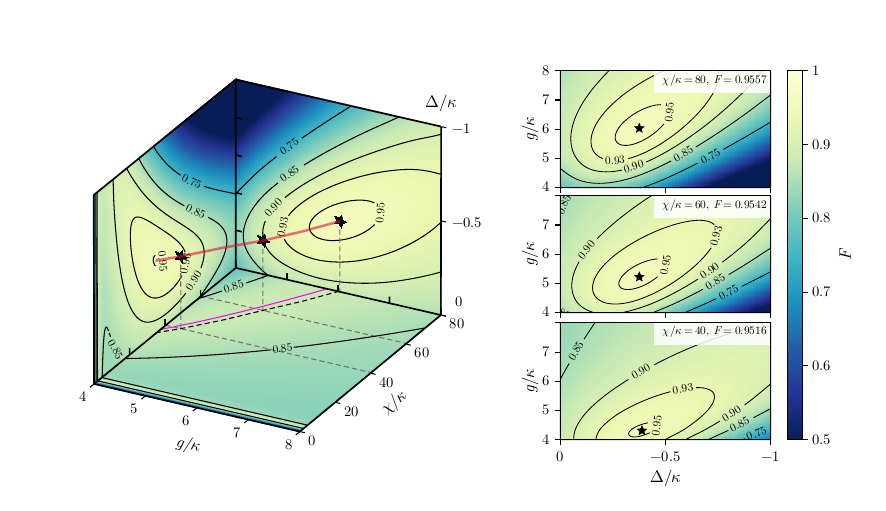}
	\caption{Optimization of $\Fent$ over $(g,\chi,\Delta)$.
    Left: $\Fent$ on three orthogonal cuts of the parameter box:
    $\Delta=0$ (floor, with the magenta ridge of the optimal $g$ at each $\chi$), $\chi/\kappa=80$ (back wall) and $g/\kappa=4$ (left wall).
    Color 
    and thin black contours 
    as in Fig.~\ref{fig:BellFidelity}.
    The red line is the profile optimum, i.e. the maximum over $g$ and $\Delta$ at each $\chi$; the stars mark $\chi/\kappa=40$, $60$ and $80$.
    Dashed black lines are the drop lines from these stars to the cut planes and the projection of red line onto the floor.
    Right: $\Fent(\Delta,g)$ at fixed $\chi/\kappa=80$, $60$, $40$; each shows a single interior maximum ($\star$), the point reached by the profile optimum.
\label{fig:BellFidelityAlgo}}
\end{figure*}

\subsection{Bell fidelity optimization}
\label{app:FidelityOptimization}

The optimization in Sec.~\ref{sec:FrequencyBinEntanglement} is carried out for the fully entangled fraction $\Fent=\tfrac12\bigl(\sqrt{\lambda_1}+\sqrt{\lambda_2}\bigr)^2$ over $(g,\chi,\Delta)$ in units of $\kappa$ (symmetric parameters $g_A=g_B$, $\chi_{AB} = \chi_{BA}$, $\Delta_A=\Delta_B$).
The structure of the objective dictates the procedure.
At fixed $\chi$, $\Fent(g,\Delta)$ has a single interior maximum, whose ridge is tilted in the $(g,\Delta)$ plane (right panels of Fig.~\ref{fig:BellFidelityAlgo}).
Along $\chi$, however, the profile $\textstyle F_{**}(\chi)=\max_{g,\Delta}\Fent(\chi, g, \Delta)$ increases monotonically and saturates (red line in Fig.~\ref{fig:BellFidelityAlgo}), with no stationary point inside the explored range $\chi/\kappa\le80$; a direct search in all three parameters would therefore stop at an essentially arbitrary point of the plateau.
We therefore treat $\chi$ as a control parameter and maximize over $(g,\Delta)$ at each fixed $\chi$ separately.
The optimal detuning drifts only weakly, from $\Delta/\kappa\simeq - 0.43$ at $\chi/\kappa=10$ to $-0.38$ at $\chi/\kappa=80$, whereas the optimal $g$ grows from $2.2\kappa$ to $6.0\kappa$; this is why the ridge in the lower plot of Fig.~\ref{fig:BellFidelity}(a), computed at fixed $\Delta/\kappa = -0.38$, lies close to the maximum over both $g$ and $\Delta$.

For each $\chi/\kappa\in\{10,15,20,\dots,80\}$, we use a derivative-free pattern search of the Hooke--Jeeves type~\cite{hooke1961,kolda2003}, started from the optimum suggested by coarse maps of $\Fent(|g|,\chi)$ at $\Delta=0$ and $\Delta/\kappa = -0.38$ ($29\times41$ grids, $3\le|g|/\kappa\le10$, $0\le\chi/\kappa\le80$).
In each iteration, eight trial points are evaluated around the current best one, with steps $(s_g,s_\Delta)=(0.5,0.2)\kappa$ along $\pm g$, $\pm\Delta$, and the four diagonals---the diagonals are needed because the ridge is tilted, and axis-only moves crawl along it.
The best trial point is accepted if it improves $\Fent$; otherwise both steps are halved.
The search stops once the steps fall below $(0.03,0.01)\kappa$, which takes $57$--$97$ function evaluations per $\chi$.
The search is restricted to $1\le|g|/\kappa\le10$ and $-2 \le \Delta/\kappa \le 1$; no optimum lies near these bounds.

Every function evaluation is the SVD described above, with window $W$,
grid step $\delta\omega=0.12\,\kappa$ and emission time $t=400/\kappa$.
The $\Fent$ values are insensitive to these choices because they depend only on the two leading Schmidt coefficients: at the optimum at $\chi/\kappa=80$, the value $\Fent=0.9557$ is unchanged in the fifth digit between $t=400/\kappa$ and $2000/\kappa$ and between $\delta\omega=0.12\,\kappa$ and $0.07\,\kappa$.
Over the coarse maps, re-evaluating $88$ points at $\delta\omega=0.08\,\kappa$ changes $\Fent$ by less than $3\times10^{-5}$ everywhere for $|g|/\kappa\ge4$ (median below $10^{-6}$); the only deviations (up to $1.4\times10^{-2}$) occur at $|g|/\kappa=3$ and $\chi/\kappa\gtrsim56$, where $\Fent<0.65$.
For this reason, Figs.~\ref{fig:BellFidelity} and~\ref{fig:BellFidelityAlgo} show only $|g|/\kappa \ge 4$. The optimum itself has $|g|/\kappa<4$ only for $\chi/\kappa\lesssim30$, where the deviations above do not occur.

\section*{References}
\bibliography{bibliography}
\end{document}